\documentclass[12pt,paper]{JHEP} 
\usepackage{graphicx}

\newcommand{\be}{\begin{equation}}
\newcommand{\ee}{\end{equation}}
\newcommand{\ba}{\begin{eqnarray}}
\newcommand{\ea}{\end{eqnarray}}

\newcommand{\im}{\mbox{Im}}

\preprint{LU TP 04-02\\
hep-ph/0401039\\
January 2004}

\title{$\pi\pi$ Scattering in Three Flavour ChPT\thanks{Supported
in part by the European Union TMR
network, Contract No. HPRN-CT-2002-00311  (EURIDICE)}
}
\vfill
\author{Johan Bijnens, Pierre Dhonte\\
Department of Theoretical Physics 2, Lund University,\\
S\"olvegatan 14A, S 223-62 Lund, Sweden}
\author{Pere Talavera\\
Departament de F\'\i sica i Enginyeria Nuclear, Universitat Polit\`ecnica
de Catalunya,\\Jordi Girona 1-3, E 08034 Barcelona, Spain}

\abstract{
We present the scattering lengths for the $\pi\pi$ processes in
the three flavour Chiral Perturbation Theory (ChPT) framework at 
next-to-next-to-leading order. We then combine this calculation with
the determination of the parameters from $K_{e4}$ and the masses
and decay constants and compare with the results of a dispersive
analysis of $\pi\pi$ scattering. The comparison indicates a small
but nonzero value for the $1/N_c$ suppressed NLO low energy constants
$L_4^r$ and $L_6^r$.
}
\keywords{Chiral Lagrangians, Spontaneous Symmetry Breaking} 

\begin{document}

\section{Introduction}

The $\pi\pi$ scattering amplitude is one of the fundamental observables in
low energy particle physics and has been since long the subject of many
studies. The pions are the lightest strongly interacting particles
and have thus a special status. Their properties are also strongly
influenced by the chiral symmetry present in the limit of massless
quarks in Quantum Chromo Dynamics (QCD).

In this paper we add one more step in the discussion of pion properties.
We calculate $\pi\pi$ scattering in three flavour Chiral Perturbation Theory
(ChPT) in order to complete the connection between the $K_{\ell4}$ form factors
and $\pi\pi$ scattering following from chiral symmetry. We stay in the isospin
limit here and neglect electromagnetic effects.
But first we give a
short historical overview of chiral symmetry relevant to $\pi\pi$ scattering.

Chiral symmetry was introduced a long time ago and used in the form of
current algebra and the PCAC assumptions (partial conservation of
axial currents.) Weinberg \cite{Weinbergpipi} used these methods to derive
a result for $\pi\pi$ scattering valid to lowest order in meson masses
and momenta. It was later realized that the assumptions of analyticity
used in many PCAC type of analyses were not always true. This allowed to
calculate often the leading nonanalytic corrections to the lowest order
PCAC results. This line of work within the PCAC methods has been reviewed
in Ref.~\cite{Pagels} where also references to earlier work can be found.
The more modern method of using chiral symmetry in the form of an effective
field theory was introduced by Weinberg \cite{Weinberg} and systematized
by Gasser and Leutwyler \cite{GL1}. This is now known as ChPT.
They performed the full
next-to-leading order (NLO) calculation of $\pi\pi$ scattering \cite{GLpipi}
as the first major application. The parameters necessary for determining the
$\pi\pi$ scattering lengths at threshold had to be taken from the $D$ wave
experimental results. A reasonable agreement with the experimental result
was obtained. Gasser and Leutwyler also extended ChPT to the three flavour
sector including the strange quark in addition to the up and down quark
in the ChPT formalism \cite{GL2}. This allowed to determine the parameters
necessary for prediction $\pi\pi$ scattering to be determined from the
absolute values of the form factors in $K_{\ell4}$ decays. Note that the phase
and the absolute values are in principle separately measurable quantities there
such that the relation between the phase as determined by the relevant $\pi\pi$
scattering phase and the absolute value is not a trivial prediction.
These NLO calculations of $K_{\ell4}$ decays were performed and again
led to a reasonable agreement with the known values
\cite{Bijnenskl4,Riggenbach}. The three flavour expression for $\pi\pi$
scattering was first calculated in \cite{BKM1} and later independently in
\cite{Knecht1}. 

The first step at 
next-to-next-to-leading order (NNLO) was done in Ref.~\cite{Knecht1} where
the dispersive part of the amplitude was determined.
The full two loop calculation in two flavour ChPT was performed
in Ref.~\cite{BCEGS1,BCEGS2}. The NNLO calculation of $K_{\ell4}$ has
also been performed \cite{ABT2,ABT3}
and was used to study $\pi\pi$ scattering via
the two flavour NNLO calculation using the relation between the NLO
order parameters in two and three flavour ChPT derived in Ref.~\cite{GL1}.
This still leaves an uncertainty since in order to have full control at
NNLO the corrections to those relations need also to be determined.
This can be done in principle by integrating out the kaons and eta degrees
of freedom out of the three flavour ChPT NNLO generating functional but
this has not been done so far. Alternatively one can directly
calculate the NNLO amplitude for $\pi\pi$ scattering in three flavour
ChPT. This is what has been done in this paper.

A different way to describe theoretically $\pi\pi$ scattering is to use
the constraints from analyticity and unitarity. These lead to
many different sum rules but especially after crossing properties
have been included the resulting set of equations becomes very constraining.
These are known as the Roy equations \cite{Roy}. They were analyzed extensively
in the seventies, see e.g. \cite{Basdevant}. The available high energy data
together with the Roy equations allowed to describe $\pi\pi$ scattering
in terms of two parameters usually chosen to be the scattering lengths
in the $S$ wave channel for isospin 0 and 2, $a_0^0$ and $a^2_0$ 
respectively.
$a_0^0$ was then essentially fixed by measuring the difference in phase of the
 $K_{e4}$ form factors. This measurement has been dominated for a long time
by the experiment of Ref.~\cite{Rosselet}. The result was in disagreement
with the Weinberg prediction \cite{Weinbergpipi} but in borderline
agreement with the NLO order one \cite{GLpipi}. However, the central values was
rather different from the ChPT prediction. If this central value turned out
to be correct, the consequences for the low energy structure of QCD would
be quite strong. In particular the chiral symmetry breaking might not be
driven by the simple quark anti-quark condensate, see \cite{Stern1}
and references therein. This was the origin for the renewed interest in
$\pi\pi$ scattering. The analysis of the Roy equations has been updated
in Ref.~\cite{ACGL}. This was then combined with the constraints from
chiral symmetry in Ref.~\cite{CGL1,CGL2}. The constraints used
were the solutions of the Roy equations of \cite{ACGL}, the ChPT two flavour
calculations of $\pi\pi$ scattering \cite{BCEGS1,BCEGS2} and the pion
scalar form factor \cite{BCT}. It also used the method of determination of
low energy constants from sum rules over $\pi\pi$ phases from
\cite{Knecht2}. This analysis led to very constrained predictions for
the scattering lengths\footnote{The conclusions of that analysis have
been challenged in \cite{Yndurain},
the reply of the authors of \cite{CGL1,CGL2} can be found in \cite{Caprini}.}.
These were confirmed nicely by the E865 experiment
at BNL \cite{Pislak1,Pislak2}. A similar analysis but without the constraint
from the pion scalar radius can be found in \cite{DFGS}.

In this paper we will compare the predictions for the $\pi\pi$ scattering
lengths in three flavour ChPT with the experimental input from $K_{\ell4}$
decays to NNLO. We find an acceptable agreement as discussed in
Sect.~\ref{numerical}.

In two flavour ChPT it is clear now that spontaneous
chiral symmetry breaking goes through the
quark anti-quark condensate \cite{CGL1,CGL2}. There is still a possibility that
the behaviour when also the strange quark mass goes through zero, is
qualitatively different. This is discussed in the recent work by Stern
and collaborators \cite{Stern2,Stern3}. The argument is that large disconnected
loop contributions from strange quarks, via kaons and etas, can be large,
making a convergent three flavour ChPT difficult to achieve in the usual
sense \cite{Stern3,LJP}. 
This has been studied in some detail in \cite{ABT3} and also
in the context of the three flavour ChPT calculations of the various scalar
form factors \cite{BD}. The masses and decay constants, see Ref.~\cite{ABT1}
and references therein, showed the possibility of this behaviour.
The various vector form factors calculated did not seem to have problems
with convergence \cite{Post1,Post2,BT1,BT2}. It turns out that this
calculation does not provide much more information than the scalar form factors
\cite{BD} did. Work is in progress to extend the $\pi K$ scattering also
to NNLO. This might allow us to shed more light on this issue.

This paper is organized as follows. In Sect.~\ref{chpt} we give a short
overview of ChPT and the references for the methods of NNLO calculations.
In Sect.~\ref{pipigeneral} the general properties of the $\pi\pi$ scattering
amplitude are described and the quantities used later
defined. Sect.~\ref{analytical} gives an overview of our main result, the
calculation of the $\pi\pi$ scattering amplitude to NNLO in three flavour ChPT.
We also present here some plots showing the importance of the various
contributions. The inputs we use to do the numerical analysis are
described in Sect.~\ref{inputs}. The main numerical analysis is
presented in Sect.~\ref{numerical} and we give our main conclusions
in Sect.~\ref{conclusions}. The appendices are devoted to
giving references
about the various loop integrals used in this work and the explicit
expressions at NNLO for $\pi\pi$ scattering.

\section{Chiral Perturbation Theory}
\label{chpt}

ChPT is the effective field theory for QCD at low energies introduced by
Weinberg, Gasser and Leutwyler \cite{Weinberg,GL1,GL2}. Introductory
lectures can be found in Ref.~\cite{chptlectures}. This leads to an expansion
in quark masses and meson momenta generically labeled $p$ and
assumes $m_q\sim p^2$ since for an on-shell meson $p_\pi^2 = m_\pi^2$.
The ChPT formalism exists both for two light flavours, up and down, referred
to as $SU(2)$ ChPT, and for three light flavours, up,
down and strange, referred
to as $SU(3)$ ChPT.
The Lagrangian for the strong and semi leptonic mesonic sector
to NNLO can be written as
\be
{\cal L} = {\cal L}_2+ {\cal L}_4+{\cal L}_6\,,
\ee
where the subscript refers to the chiral order.
The lowest order Lagrangian is
\be
\label{L2}
{\cal L}_{2} = \frac{F_0^2}{4} \langle u_\mu u^\mu + \chi_+ \rangle \,.
\ee
The mesonic fields enter via
\be
u = \exp\left(\frac{i M}{F_0 \sqrt{2}}\right)\,,\quad
M =\left(\begin{array}{ccc}
\frac{1}{\sqrt{2}}\pi^0+\frac{1}{\sqrt{6}}\eta & \pi^+ & K^+\\
\pi^- & \frac{-1}{\sqrt{2}}\pi^0+\frac{1}{\sqrt{6}}\eta & K^0\\
K^- & \overline{K^0} & \frac{-2}{\sqrt{6}}\eta
         \end{array}\right)\,
\ee
and the quantity $u_\mu$ also contains the external vector ($v_\mu$)
and axial-vector ($a_\mu$)
currents
\be
\label{covariant}
u_\mu = i 
(u^\dagger \partial_\mu u - \partial_\mu u u^\dagger 
 -i u^\dagger r_\mu u + i u l_\mu u^\dagger)\,,
\quad
l_\mu(r_\mu) = v_\mu -(+) a_\mu\,.
\ee
The scalar ($s$) and pseudo scalar ($p$) currents are contained in
\be
\chi_\pm = u^\dagger \chi u^\dagger \pm u \chi^\dagger u\,,\quad
\chi = 2 B_0\left(s+ip\right)\,.
\ee
The $p^4$ or NLO Lagrangian, ${\cal L}_4$, was introduced in Ref.~\cite{GL2}
and reads
\ba
\label{L4}
{\cal L}_{4}&=&
\hspace{0.34cm} L_1 \langle u_\mu u^\mu \rangle^2
+L_2 \langle u_\mu u_\nu  \rangle
     \langle u^\mu u^\nu  \rangle  \nonumber\\&& 
+L_3 \langle u^\mu u_\mu u^\nu u_\nu \rangle 
+L_4 \langle u^\mu  u_\mu  \rangle \langle \chi_+ \rangle \nonumber\\&&
+L_5 \langle u^\mu u_\mu \chi_+ \rangle
+L_6 \langle \chi_+ \rangle^2 \nonumber\\&&
+L_7 \langle \chi_- \rangle^2
+L_8 \langle \chi_+ \chi_- \rangle \nonumber\\&&
\hspace{-0.14cm}
-i L_9 \langle F^R_{\mu\nu} u u^\mu u^\nu u^\dagger 
             + F^L_{\mu\nu} u^\dagger u^\mu u^\nu u \rangle 
+L_{10} \langle   F^R_{\mu\nu}  U F^{L\mu\nu} U^\dagger 
\rangle 
\nonumber\\&&
+H_1 \langle F_{\mu\nu}^RF^{\mu\nu R} + F_{\mu\nu}^LF^{\mu\nu L} \rangle
+H_2 \langle \chi_+^2 - \chi_-^2 \rangle /4 
\,.
\ea
 The $L_9$ and $L_{10}$ terms introduce also the field strength tensor
\be
F_{\mu\nu}^{L(R)} = \partial_\mu l(r)_\nu -\partial_\nu l(r)_\mu -i 
\left[ l(r)_\mu , l(r)_\nu \right]\,.
\ee
The two terms proportional to
$H_1$ and $H_2$ are high energy contact terms and are not involved in 
physical amplitudes,
$L_9$ and $L_{10}$ play only a minor role for the quantities discussed
in this paper.

We quote the schematic form of the NNLO Lagrangian in the three flavour case
\be
\label{L6}
{\cal L}_6 = \sum_{i=1,94} C_i\,O_i
\ee
and refer to \cite{BCE} for their explicit expressions.
The last four terms are contact terms~\cite{BCE}.

The ultra-violet divergences produced by loop diagrams of order
$p^4$ and $p^6$ cancel in the process of renormalization with the divergences
extracted from the low energy constants $L_i$'s and $C_i$'s.
We use dimensional regularization and
the modified minimal subtraction
$(\overline{MS})$ version usually used in ChPT.
An extensive description of the
regularization and renormalization procedure including
the freedom involved can be found in Refs.~\cite{BCEGS2} and \cite{BCE2}.

The subtraction of divergences is done explicitly by 
\be 
L_i = (c\mu)^{d-4}[\hat{\Gamma}_i \Lambda + L_i^r(\mu)]
\ee
and
\be
C_i = \frac{(c\mu)^{2(d-4)}}{F^2}\left[C_i^r(\mu)-\left(\Gamma_i^{(1)}
+\Gamma_i^{(L)}(\mu)\right) \Lambda-\Gamma_i^{(2)} \Lambda^2\right]
\ee
where $c$ and $\Lambda$ are defined by 
\be 
\ln{c} = -\frac{1}{2}\left[\ln{4\pi}+\Gamma^\prime(1)+1\right]\,,
\ee
\be 
\Lambda = \frac{1}{16\pi^2(d-4)}\,.
\ee
The coefficients $\hat{\Gamma}_i$, $\Gamma_i^{(1)}$ and $\Gamma_i^{(2)}$
are constants while the $\Gamma_i^{(L)}$\,'s are linear combinations of
the $L_i^r(\mu)$\,'s.
Their explicit expressions can all be found in \cite{BCE2} where they have been
calculated in general. 
The NLO divergences were first calculated in
Ref.~\cite{GL1,GL2}
and the doubles poles at NNLO first in Ref.~\cite{BCE3}.

\section{The $\pi\pi$ amplitude: general properties}
\label{pipigeneral}

The $\pi\pi$ scattering amplitude in all the relevant channels can be written
as a function $A(s,t,u)$ which is symmetric in the last two
arguments:
\be
A\left(\pi^a(p_1)\pi^b(p_2)\to\pi^c(p_3)\pi^d(p_4)\right)
= \delta^{ab}\delta^{cd} A(s,t,u)
+ \delta^{ac}\delta^{bd} A(t,u,s)
+ \delta^{ad}\delta^{bc} A(u,t,s)\,.
\ee
$s,t,u$ are the usual Mandelstam variables
\be
s = (p_1+p_2)^2\,\quad t=(p_1-p_3)^2\quad\mbox{and}\quad u=(p_1-p_4)^2\,.
\ee

The various isospin amplitudes $T^I$ can be written
in terms of this function as
\ba
T^0(s,t) &=& 3 A(s,t,u) + A(t,u,s) +A(u,s,t)\,,
\nonumber\\
T^1(s,t) &=& A(t,u,s) - A(u,s,t)\,,
\nonumber\\
T^2(s,t) &=& A(t,u,s) +A(u,s,t)\,,
\ea
where the kinematical variables $t,u$ can be expressed in terms of $s$ 
and $\cos\theta$ as
\be
t = -\frac{1}{2}(s-4m_\pi^2)(1-\cos\theta)\,,\quad
u = -\frac{1}{2}(s-4m_\pi^2)(1+\cos\theta)\,.
\ee

The various amplitudes can be expanded in partial waves via
\be
T^I(s,t) = 
32\pi\sum_{\ell=0}^\infty (2\ell+1) P_\ell(\cos\theta) t^I_\ell(s)\,.
\ee
Near threshold these are expanded in terms of the threshold parameters
\be
\label{defaij}
t^I_\ell = q^{2\ell}\left(a^I_\ell + b^I_\ell q^2 + {\cal O}(q^4)\right)\,,
\quad
q^2 = \frac{1}{4}\left(s-4 m_\pi^2\right)\,.
\ee

Below the inelastic threshold the partial waves satisfy
\be
\im t^I_\ell(s) = \sigma(s) \left|t^I_\ell(s)\right|^2,\quad\quad
 \sigma(s) = \sqrt{1-\frac{4 m_\pi^2}{s}}\,.
\ee
In this regime the partial waves can be written in terms of the phase-shifts
as
\be
t^I_\ell(s) = \frac{1}{\sqrt{1-(4m_\pi^2/s)}}\,\frac{1}{2i}\left\{
e^{2i\delta^I_\ell(s)}-1\right\}\,.
\ee

In ChPT the inelasticity only starts at order $p^8$. 
The $p^2$ result
has only nonzero items for $t^0_0$, $t^1_1$ and $t^2_0$. As a consequence
the imaginary parts for all other partial waves starts only at order $p^8$.
In \cite{Stern1} it has been shown that thus up to order $p^8$ the amplitude
can be written as
\ba
\label{defVi}
A(s,t,u) &=& C(s,t,u) +32\pi\Bigg(\frac{1}{3}V^0(s)
+\frac{3}{2}(s-u)V^1(t)+\frac{3}{2}(s-t)V^1(u)
\nonumber\\&&
+\frac{1}{2}V^2(t)+\frac{1}{2}V^2(u)-\frac{1}{3}V^2(s)\Bigg)
\,.
\ea
The function $V^I(s)$ have a polynomial ambiguity but we have
chosen to keep it in this more general form. The $V^I(s)$ contain
the singularities in the $\pi\pi$ amplitudes from intermediate states
with isospin $I$ in the various channels. They obey the relations
\be
\im V^{0,2}(s) = \im t^{0,2}_0(s)\,,\quad\quad
\im V^1(s) =\frac{1}{s-4 m_\pi^2} \im t^1_1(s)\,.
\ee
The polynomial $C(s,t,u)$ can be written using $s+t+u=4m_\pi^2$ in
the form
\be
\label{defci}
C(s,t,u) = c_1+c_2 s + c_3 s^2 +c_4(t-u)^2+c_5 s^3 + c_6 s(t-u)^2\,.
\ee

\section{ChPT results}
\label{analytical}

\subsection{Two Flavour ChPT}

The lowest order was derived by Weinberg \cite{Weinbergpipi} a long time
ago and corresponds to
\be
\label{pipiLO}
c_1 = \frac{m_\pi^2}{F_\pi^2}\,,\quad\quad
c_2 = -\frac{1}{F_\pi^2}\,,
\ee
in (\ref{defci}).
The next order was derived by Gasser and Leutwyler \cite{GLpipi} and
the full calculation of order $p^6$ was performed in \cite{BCEGS1,BCEGS2}.

\subsection{Three Flavour ChPT}

The lowest order is identical to the two-flavour case of Eq.~(\ref{pipiLO}).
The order $p^4$  $\pi\pi$ scattering amplitude in three flavour ChPT was
first published
in \cite{BKM1} App.~A. It can also be found in \cite{Knecht1}. Notice that
the contribution from $L_1^r$ and $L_3^r$ is missing in \cite{BKM1}. Our
result at this order is in full agreement with the corrected version. 
We have expressed the result
in terms of the functions defined in Eq.~( \ref{defVi}) in App.~\ref{appB}.

The $p^6$ expressions we present are those corresponding to the $p^4$
results expressed in the physical masses and decay constants.
The order $p^6$ expression is our main result. Expressed in the
polynomial $C(s,t,u)$ and the functions $V^{0,1,2}(s)$ it is shown
in App.~\ref{appB}.

\subsection{A first numerical look}
\label{firstnum}

In this subsection we present a first look at the numerical results
for the two loop amplitudes. We choose as input the pion decay constant,
the charged pion mass, an averaged kaon mass with electromagnetic effects
removed and the physical eta mass.
\ba
F_\pi &=& 92.4~\mbox{MeV}\,,\quad m_\pi = m_{\pi^+} = 139.56995~\mbox{MeV}\,,
\nonumber\\
m_K &=& 494.53~\mbox{MeV}  \,,\quad m_\eta = 547.3~\mbox{MeV}\,.
\ea
The subtraction scale $\mu = 770~\mbox{MeV}$ is used throughout
the paper unless otherwise mentioned explicitly.

We first compare the $SU(2)$ and $SU(3)$ ChPT results for the three main
partial waves for all low energy constants set to zero at the scale 770~MeV.
So we set $L_i^r = C_i^r = l_i^r = c_i^r = 0$ and compare
the pure loop results for the $SU(2)$ and $SU(3)$ ChPT.

\FIGURE{
\includegraphics[angle=270,width=0.75\textwidth]{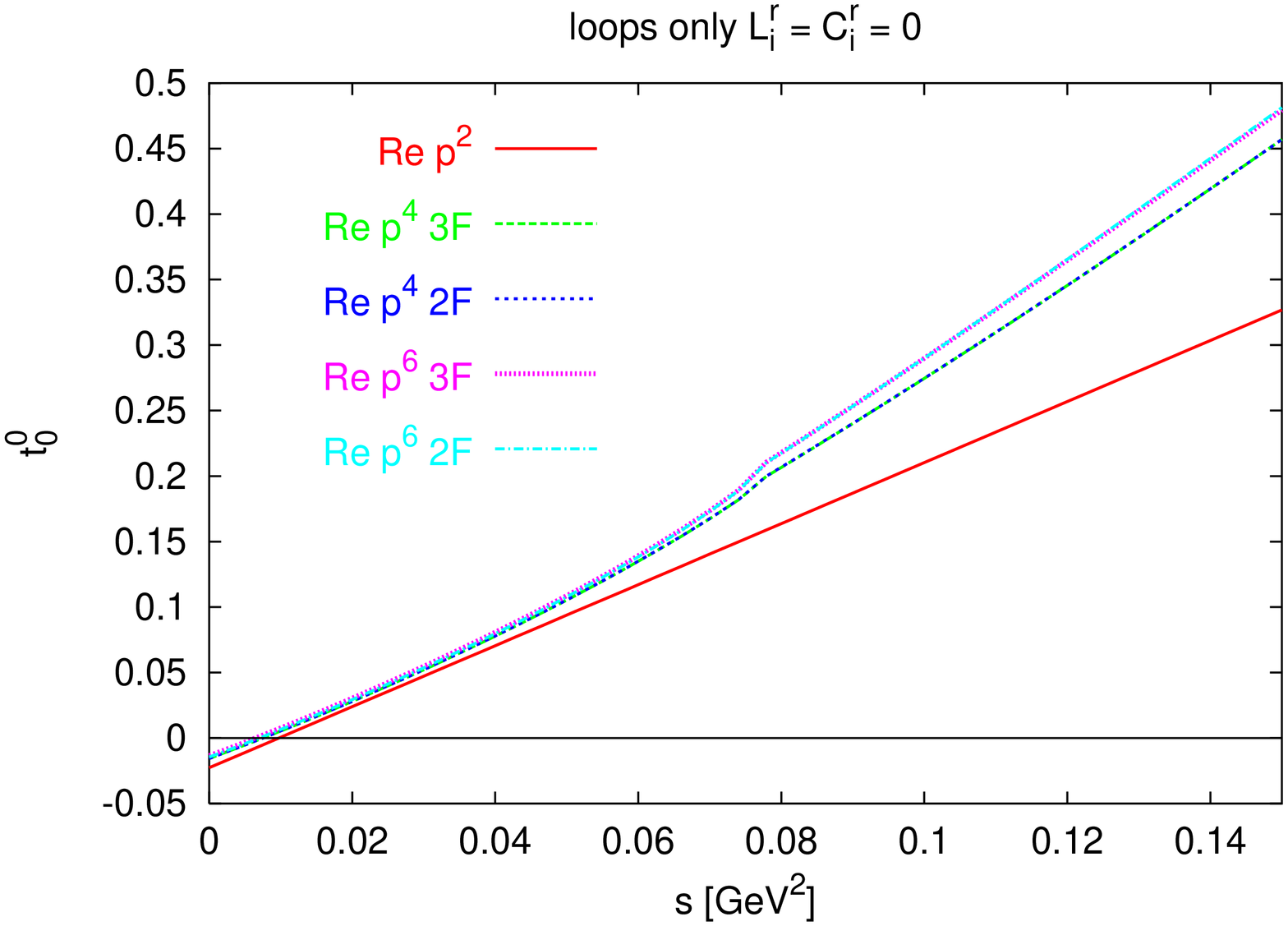}
\caption{\label{figt00}  Results for the $t^0_0$ partial wave:
The $SU(2)$ ChPT results of \cite{BCEGS2}
for all LECs equal to zero, labeled 2F, compared to the results
for the $SU(3)$ ChPT results also for all LECs equal to zero at 
$\mu= 770~\mbox{MeV}$, labeled 3F.}
}

In Fig.~\ref{figt00} we have plotted the partial wave amplitude $t^0_0(s)$
Notice the extremely small difference between the two results, showing that
this channel is obviously dominated by the pion loops and the kaon and eta
have only a fairly small effect.

\FIGURE{
\includegraphics[angle=270,width=0.75\textwidth]{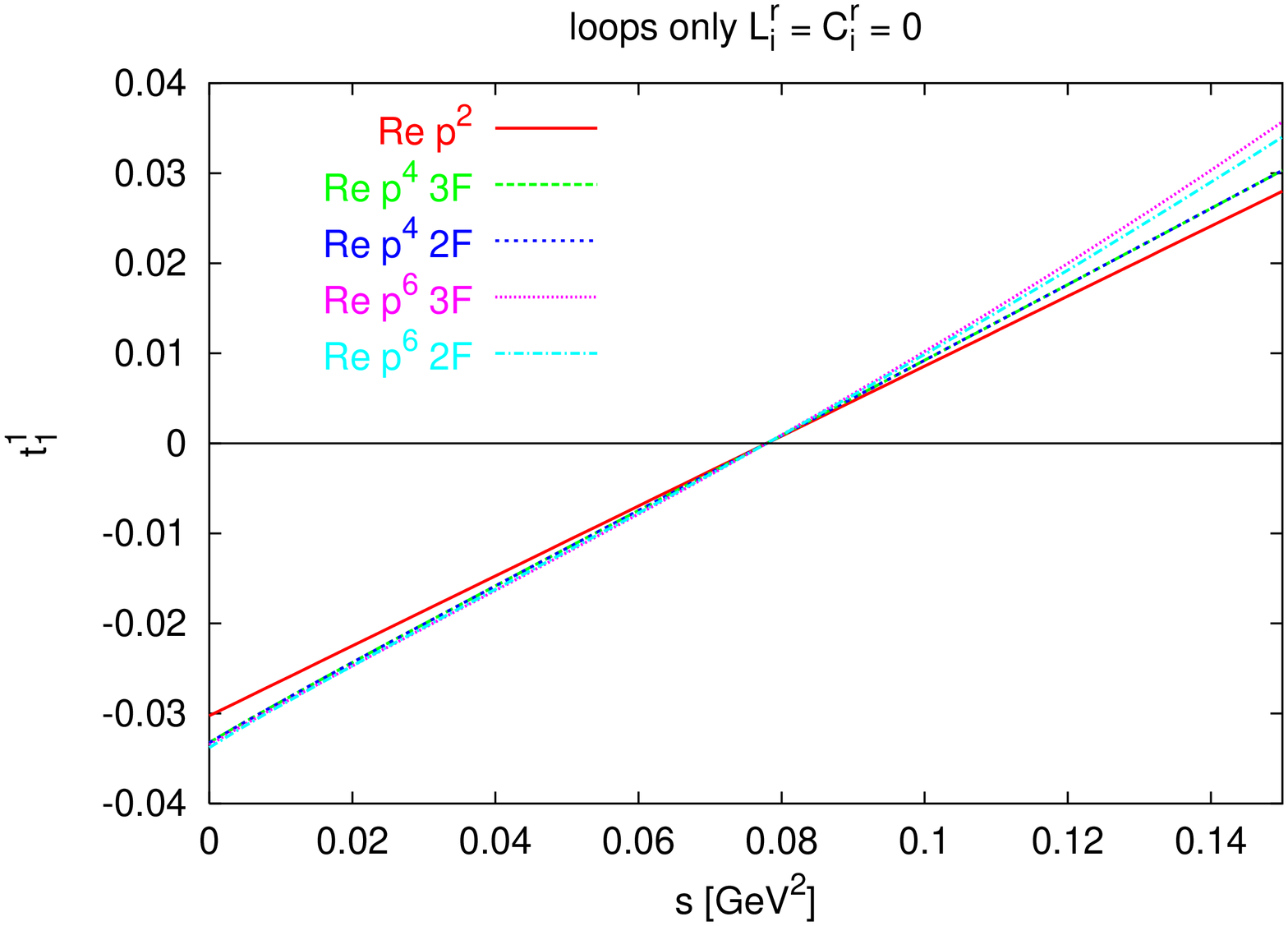}

\caption{\label{figt11} Results for the $t^1_1$ partial wave:
The $SU(2)$ ChPT results of \cite{BCEGS2}
for all LECs equal to zero, labeled 2F, compared to the results
for the $SU(3)$ ChPT results also for all LECs equal to zero at 
$\mu= 770~\mbox{MeV}$, labeled 3F.}
}

For the $t^1_1$ partial wave, shown in Fig.~\ref{figt11}
the order $p^4$ results are very similar
in both cases again but there is somewhat more difference in the
order $p^6$ contributions, this channel has a much stronger effect from the
LECs, see below, so this difference does not play much of a role.

\FIGURE{
\includegraphics[angle=270,width=0.75\textwidth]{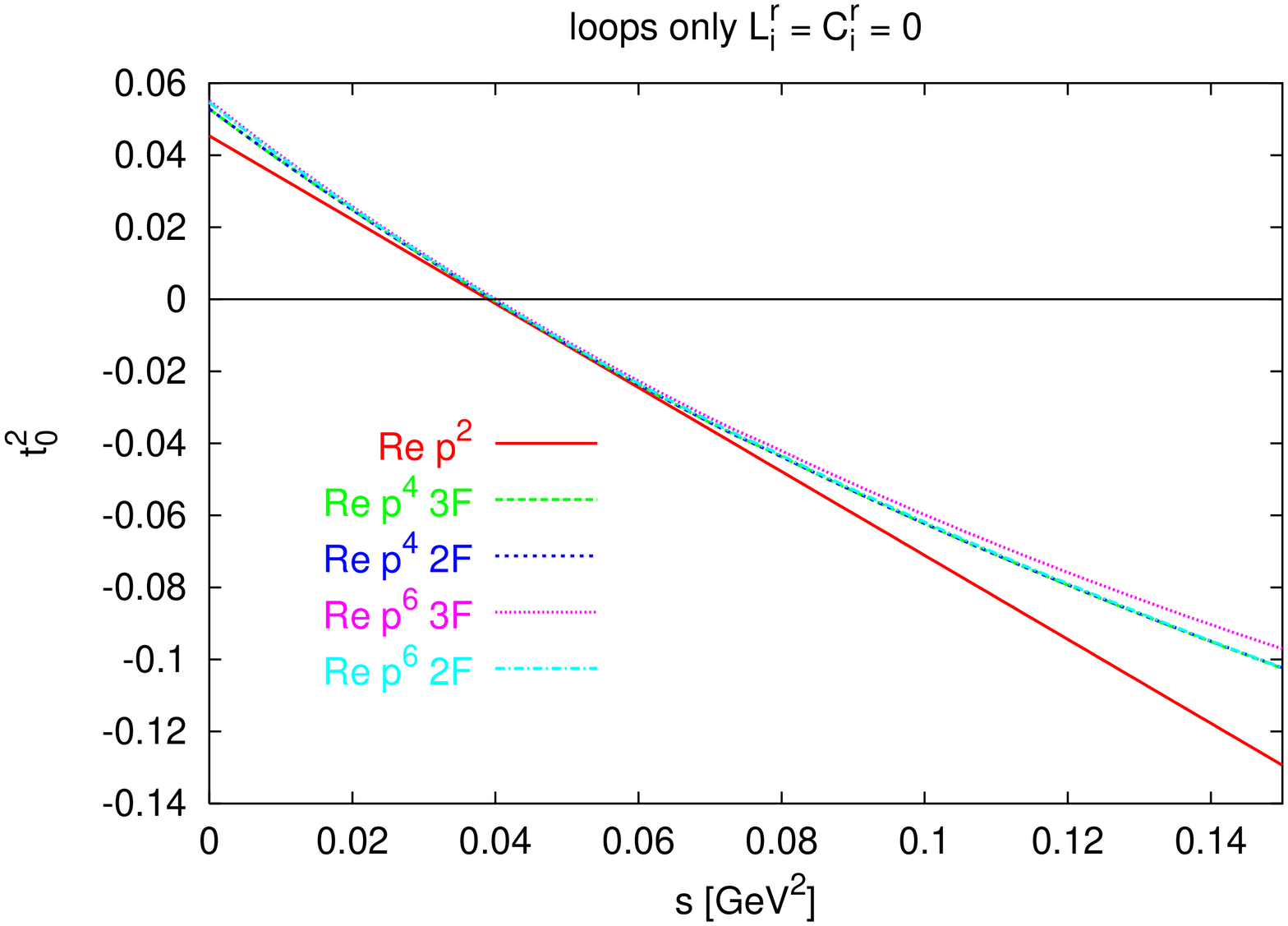}

\caption{\label{figt20} Results for the $t^2_0$ partial wave:
The $SU(2)$ ChPT results of \cite{BCEGS2}
for all LECs equal to zero, labeled 2F, compared to the results
for the $SU(3)$ ChPT results also for all LECs equal to zero at 
$\mu= 770~\mbox{MeV}$, labeled 3F.}
}

The $t^2_0$ amplitude, shown in Fig.~\ref{figt20} has a somewhat more
surprising difference. Again the order $p^4$ results for the $SU(2)$
and $SU(3)$ cases are very similar but the $p^6$ pure loop correction
is rather different. We will come back to this later.

That the difference at order $p^4$ would be small was of course expected.
In \cite{BCEGS2} the contribution from the $p^4$ loops from kaons and etas
was estimated and found to be very small.

\section{Solution of the Roy Equations and Other Inputs}
\label{inputs}

For the experimental values of the $\pi\pi$ scattering phase-shifts
we use the results from the extensive analysis of the Roy equations
done in \cite{ACGL}. This analysis has been checked with somewhat relaxed
input assumptions in \cite{DFGS}. These equations and the input
are constrained in such a way that the $\pi\pi$ phase-shifts are
determined as a function of two input parameters chosen to
be the scattering lengths $a^0_0$ and $a^2_0$.

The solution of the Roy equations has
been used in \cite{CGL1,CGL2} 
together with the constraints from the $SU(2)$ ChPT $p^6$ expression
to obtain a precise prediction of $a^0_0$ and $a^2_0$.
The input used there was the pion scalar radius evaluated using dispersive
methods and the estimates of the $p^6$ low energy constants of \cite{BCEGS2},
together with the $SU(2)$ order $p^6$ calculations of $\pi\pi$ scattering
\cite{BCEGS1,BCEGS2} and the pion scalar radius \cite{BCT}.

The conclusions of that analysis have been challenged in \cite{Yndurain},
the reply of the authors of \cite{CGL1,CGL2} can be found in \cite{Caprini}.

For our $SU(3)$ ChPT results we use as inputs the masses and decay constants
given in Sect.~\ref{firstnum} and a subtraction constant $\mu = 770$~MeV.
We work in the limit of exact isospin.

It should be mentioned that these numbers were obtained using a general refit
of all the other $L_i^r$ with a fixed $L_4^r$ and $L_6^r$ as input. The inputs
were the absolute values of the $K_{\ell4}$ form factors as measured
by the E865 experiment \cite{Pislak1,Pislak2}, decay constants and meson
masses. The fits correspond precisely to those of fit 10 in \cite{ABT4}
but with different input values for $L_4^r$ and $L_6^r$.

For the constants at order $p^6$ we need to use input values using
various estimates. The method used was proposed at NLO in
Ref.~\cite{Ecker1,Ecker2} and references therein. We do not estimate
here the NLO constants this way but only NNLO. The places where comparisons
with experiment are available are in general reasonable agreement with these
estimates. The estimates are obtained by including the main resonance exchange
contributions and putting the $p^6$ part of these amplitudes equal to
the contribution from the $C_i^r$. This procedure is obviously subtraction
point dependent and is normally only performed to leading order in the
expansion in $1/N_c$, with
$N_c$ the number of colours. Alternative approaches
exist but we will not discuss them here. Recent papers
addressing this type of issues are \cite{Pich,BGLP,MHA} and
references therein. A systematic study of this issue is clearly
important, for our present purpose the estimates seem sufficient.

The estimates from resonance exchange for the masses and decay constants
are the most uncertain. These are discussed in \cite{ABT1} and \cite{Pich}.
For the numerical results used here they have been put to zero, the naive
size estimate of \cite{ABT1} led to extremely large NNLO corrections.
The estimates of the $K_{\ell4}$ amplitudes can be found in
\cite{ABT3} after the work of \cite{BCG}. The effect of varying
these was studied in \cite{ABT3} and found to be reasonable.

The estimates of the $p^6$ contributions to $\pi\pi$ scattering we use
are those of \cite{BCEGS2}. These lead to the
contributions to the various threshold parameters given in
Table \ref{tab:thresh1}. The uncertainty on these is quite considerable
but probably within a factor of two, this is also discussed in
Ref.~\cite{CGL2}. Similar resonance estimates of $\pi\pi$ scattering can be
found in \cite{BKM2}.

In order to be able to perform a study of the dependence on $L_4^r$
and $L_6^r$ we have redone the fits to the masses, decay constants and
the $K_{\ell4}$ absolute values of the form factors with a range of
values for $L_4^r$ and $L_6^r$. This is similar to the part discussed in
\cite{ABT3} but now with the newer experimental input \cite{Pislak1,Pislak2}
included. The fits correspond exactly to the fit labeled fit 10 in
\cite{ABT4} but with various values of $L_4^r$ and $L_6^r$ as input.
These were already used in Ref.~\cite{BD} to compare with the scalar form
factors. There a general 
preference was found for the region $L_6^r \approx L_4^r-0.0003$. In that
region the corrections to the pion scalar form factor at zero were fairly
small as well as a good agreement with the pion scalar radius was obtained.
It should be kept in mind that all the other $L_i^r$ are varied together
with $L_4^r$ and $L_6^r$ in order to fit the mentioned quantities.
Reasonable fits were obtained for most values of $L_4^r$ and $L_6^r$.
Varying $L_4^r$ and $L_6^r$ {\em without} the correlated changes in the other
$L_i^r$ would lead to much larger variations than the ones shown below.

\section{Numerical Analysis}
\label{numerical}

\TABLE[t]{
\begin{tabular}{|r|r|c|c|c|r|}
\hline
         & $p^2$  & $p^4,~L_i^r=0$ & $p^6,~L_i^r=C_i^r=0$ & $p^6,~C_i^r$ only
 & Ref.~\cite{CGL2}\\
\hline
$a^0_0$  & 0.159 & 0.041 & 0.011 & 0.001 & $0.220\pm0.005$ \\
$b^0_0$  & 0.182 & 0.075 & 0.016 & 0.004 & $0.276\pm0.006$\\
\hline
$10\,a^2_0$ & $-$0.454 & 0.037 & 0.015 & $-$0.004 & $-0.444\pm0.010$\\
$10\,b^2_0$ & $-$0.908 & 0.144 & 0.029 & $-$0.014 & $-0.803\pm0.012$\\
\hline
$10\,a^1_1$ & 0.303 & 0.022 & 0.025 & 0.000 & $0.379\pm0.005$\\
$10\,b^1_1$ & $-$  & 0.005 & 0.033  & 0.000 & $ 0.057\pm0.001$\\
\hline
$10^2\,a^0_2$ & $-$ & 0.121 & 0.050 & 0.001 & $0.175\pm0.003$\\
$10^2\,b^0_2$ & $-$ & $-$0.040 & $-$0.005 &  0.005 & $-0.036\pm0.002$\\
\hline
$10^3\,a^2_2$ & $-$ & 0.492 & $-$0.187 & $-$0.003 & $0.170\pm0.013$\\
$10^3\,b^2_2$ & $-$ & $-$0.234 & $-$0.136 & $-$0.045 & $-0.326\pm0.012$\\
\hline
$10^4\,a^1_3$ & $-$ & 0.20 & 0.27 & 0.15 & $0.560\pm0.019$\\
$10^4\,b^1_3$ & $-$ & $-$0.15 & $-$0.22 & $-$ & $-0.402\pm0.018$\\
\hline
\end{tabular}
\caption{\label{tab:thresh1} The various contributions to the threshold
parameters defined in (\ref{defaij}) with the $L_i^r=0$ at $\mu=770~$MeV.
The $C_i^r$ contribution is shown separately using the estimates of
\cite{BCEGS2}. The threshold parameters are given in the corresponding power
of $m_{\pi^+}^2$. Note that we have given $a^I_\ell$ and $b^I_\ell$ always
with the same
power of ten.}
}

In order to check convergence let us first look at the various contributions
with all the low energy constants set to zero at a scale $\mu=770$~MeV.
These are shown in Table~\ref{tab:thresh1}. The angular integrals have been
performed using both a 5 point and a 8 point Gaussian integration
over $\cos\theta$. In addition the fits were performed numerically
over a range of $q^2$ above threshold. The numerical errors on the slopes
$b^I_\ell$ for $\ell=2,3$ are of the order of the last digit shown. For all
others this error is below the accuracy given. For comparison we have also
given the results of Ref.~\cite{CGL2} in the last column.

We will now compare with the
full analysis of $\pi\pi$ scattering performed with the use of the Roy
equations and the $SU(2)$ ChPT results of \cite{CGL1,CGL2}.
In principle we could redo this work with the $SU(3)$ ChPT results as
constraints instead. We have chosen not to do so, postponing a possible
more detailed comparison till after the inclusion of $\pi K$ scattering
results. Since the $SU(3)$ ChPT results contain the pion loops
which are the main effects of the $SU(2)$ ChPT calculations and
the $SU(3)$ expressions must reduce to the $SU(2)$ expressions in the
limit of a large kaon and eta mass and thus satisfy the $SU(2)$ chiral
constraints we do not expect such an analysis to differ substantially
from the one performed in Ref.~\cite{CGL1,CGL2}. 

It can already be seen from Table~\ref{tab:thresh1}
that the lowest order result together with the
pure loop contributions only already give quite a good description of the
various scattering lengths. The effects of including nonzero values for the
low energy constants should explain the difference. Notice that for
almost all cases the estimated contributions from the $p^6$ constants $C_i^r$
is rather small. The main effect is thus from the $p^4$ constants $L_i^r$.
We will now study the effects of these when they were fitted to
other data as described above with fixed values of $L_4^r$ and $L_6^r$
as input.

\FIGURE[t]{
\begin{minipage}{0.48\textwidth}
\includegraphics[angle=270,width=0.99\textwidth]{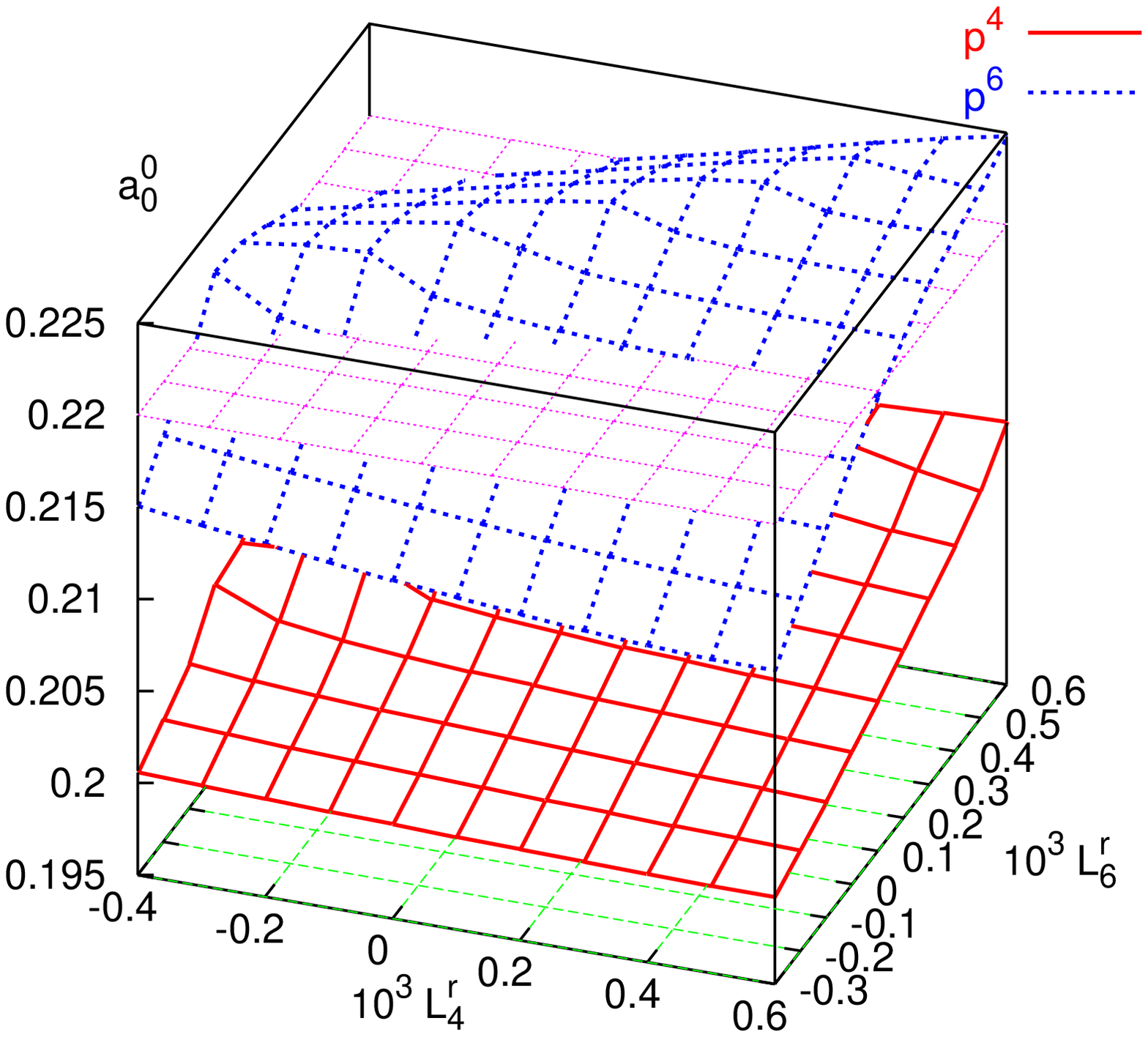}
\centerline{(a)}
\end{minipage}
\hfill
\begin{minipage}{0.48\textwidth}
\includegraphics[angle=270,width=0.99\textwidth]{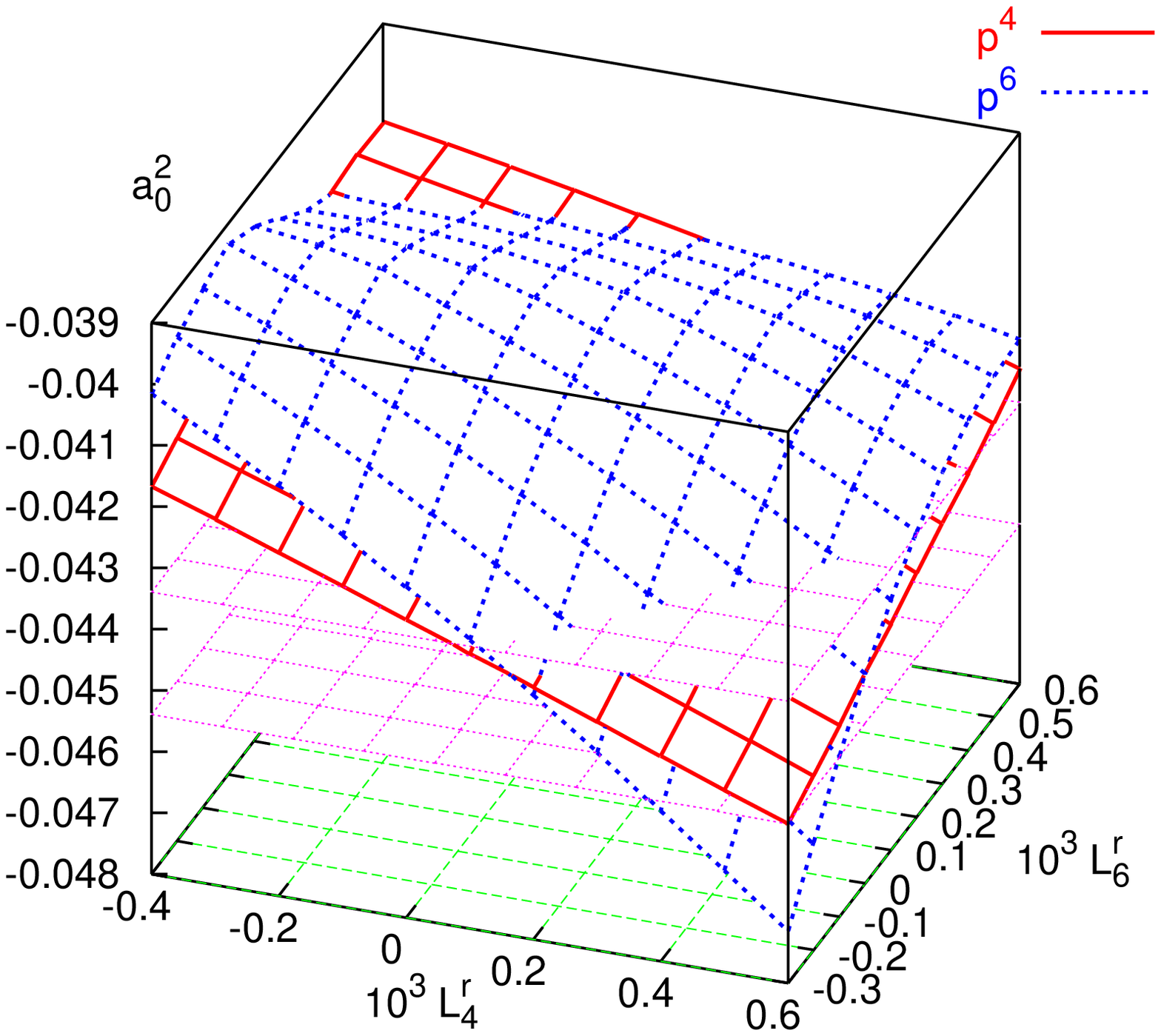}
\centerline{(b)}
\end{minipage}
\caption{\label{figa00} The scattering lengths $a^0_0$ and $a^2_0$ as
a function of the input values $L_4^r$ and $L_6^r$ with the other $L_i^r$
simultaneously refitted to the $K_{e4}$ form factors.
(a) $a^0_0$ calculated to order $p^4$ and $p^6$. The central value
of \cite{CGL2} of 0.220 is also shown. (b) $a^2_0$ at order $p^4$ and 
$p^6$. The two horizontal planes indicate the allowed region obtained
in \cite{CGL2}.}
}

In Fig.~\ref{figa00}(a) we have plotted the result for $a^0_0$ as a function
of the input values of $L_4^r$ and $L_6^r$. The lowest order value,
\be
\left.a^0_0\right|_{p^2} = 0.159\,,
\ee
is not shown on the plot. The convergence of the series is very good
and of similar quality as the two flavour result. Taking the result of
\cite{CGL2}, 
\be
\label{CGLa00}
\left.a^0_0\right|_{\cite{CGL2}} = 0.220\pm0.005\,,
\ee
we see that the agreement is excellent and no new information on $L_4^r$
and $L_6^r$ is available from this source, this also confirms the prediction
for $a^0_0$.

The result for $a^2_0$ is plotted in Fig.~\ref{figa00}(b). The lowest order
value
\be
\left.a^2_0\right|_{p^2} = -0.0454\,,
\ee
is not plotted. The series converges well over most of the $L_4^r$ and
$L_6^r$ range considered. The result should be compared with the
result \cite{CGL2}, 
\be
\label{CGLa20}
\left.a^2_0\right|_{\cite{CGL2}} = -0.0444\pm0.0010\,,
\ee
The two planes in Fig.~\ref{figa00}(b) are the error boundaries
of Eq.~(\ref{CGLa20}). We see that in order to get agreement we need to
go to the front right part of the graph. The point with $L_4^r=L_6^r=0$
is outside the error band and has the value
\be
\left.a^2_0\right|_{L_4^r=L_6^r=0} = -0.0410\,.
\ee
If we take a closer look at the scalar radius results of \cite{BD} and
especially at Fig.~11(a) there, we see that the dispersive and the $SU(3)$
ChPT result for the scalar radius are in good agreement
at\footnote{In \cite{BD} in addition a small correction to the form factor
at zero was required.}
\be
\label{L4L6radius}
L_6^r \approx L_4^4-0.0004\,.
\ee
The value of the scalar form factor there is in good agreement with the
result used in \cite{CGL1,CGL2} of $<r^2>^S_\pi= 0.61\pm0.04$~fm$^2$.
Following the line (\ref{L4L6radius}) in our results leads to a virtually
constant prediction of
\be
\label{a20result}
a^2_0 = -0.0433\,.
\ee
This is in reasonable agreement with the result of \cite{CGL2}. The $SU(3)$
ChPT result thus confirms the result of Ref.~\cite{CGL2} when the
constraint for the scalar radius is taken into account. We have not performed
a full error analysis for the result (\ref{a20result}) similar to the
one performed in Ref.~\cite{ABT3}, but we expect the errors coming from the
various uncertainties on the $L_i^r$ to be similar to the ones quoted there.

At this point we have checked the agreement for the two main input
parameters for the dispersive calculations. How well do the other
threshold parameters compare? We can show plots similar to the ones for
$a^0_0$ and $a^2_0$ shown in Fig.~\ref{figa00}, but they do not 
provide any essential new information.
We have first given in Table ~\ref{tab:thresh2} the results for the
various threshold parameters for the input values from fit 10 and fits A,B,C
as defined in Ref.~\cite{BD}. This table can be seen as the extension of the
one given in that reference for the masses, decay constants and scalar radius
to the $\pi\pi$ scattering threshold parameters.

\TABLE[t]{
\begin{tabular}{|c|c|c|c|c|c|c|c|}
\hline
      &   &\multicolumn{3}{c|}{ fit 10} & fit A & fit B & fit C \\
\hline
         & $p^2$ & $p^4$ & $p^6$ & total & total & total & total \\
\hline  
$a^0_0$  & 0.159 & 0.044 & 0.016 & 0.219 & 0.220 & 0.220 & 0.221\\
$b^0_0$  & 0.182 & 0.073 & 0.025 & 0.279 & 0.282 &0.282 &0.282 \\
\hline
$10\,a^2_0$ & $-$0.454 & 0.030 & 0.013 & $-$0.410 &$-$0.427&$-$0.433&$-$0.428\\
$10\,b^2_0$ & $-$0.908 & 0.151 & 0.025 & $-$0.731  &$-$0.755&$-$0.761&$-$0.760\\\hline
$10\,a^1_1$ & 0.303 & 0.052  & 0.031 & 0.385 &0.388&0.389&0.389\\
$10\,b^1_1$ & $-$  & 0.029 & 0.038 & 0.067 &0.064&0.063&0.063\\
\hline
$10^2\,a^0_2$ & $-$ & 0.153 & 0.080 &0.233 &0.223&0.220&0.221\\
$10^2\,b^0_2$ & $-$ &$-$0.040 & 0.007 & $-$0.033  &$-$0.035&$-$0.036&$-$0.036\\
\hline
$10^3\,a^2_2$ & $-$ & 0.327  &$-$0.106  &0.221  &0.219&0.218&0.221\\
$10^3\,b^2_2$ & $-$ &$-$0.234  & $-$0.151  &$-$0.385  &$-$0.386&$-$0.385&$-$0.387\\
\hline
$10^4\,a^1_3$ & $-$ & 0.20 & 0.44  &0.64  &0.62&0.62&0.62\\
$10^4\,b^1_3$ & $-$ & $-$0.15 & $-$0.20  & $-$0.35 &$-$0.34&$-$0.34&$-$0.34\\
\hline
\end{tabular}
\caption{\label{tab:thresh2} The values of the threshold
parameters defined in (\ref{defaij}) for the values of the input
parameters for the fits 10~\cite{ABT4} and A,B,C~\cite{BD}. The lowest order
values and the contributions from the $C_i^r$ as given in
Table~\ref{tab:thresh1} are included.
The threshold parameters are given in the corresponding power
of $m_{\pi^+}^2$. Note that we have given $a^I_\ell$ and $b^I_\ell$ always
with the same
power of ten. For fit 10 we also give the three orders separately.}
}

As can be seen a reasonable agreement is obtained for most threshold
parameters studied. We have not performed a full error analysis but
a first estimate is about half the $p^6$ contribution plus the size
of the estimated contribution from the $C_i^r$. Only for $b^2_0$ there is
a mild discrepancy with these criteria.

A possible further test can be done by comparing parameters relevant in
the subthreshold expansion. Here often a faster convergence of the
chiral series is expected. Examples of such parameters are the coefficients
$c_i$ of the polynomial $C(s,t,u)$ defined in Eq.~(\ref{defci}) after
the polynomial ambiguity in the $V^I(t)$ is removed by requiring
\be
\frac{d^n}{ds^n} V^I(s) = 0  \quad 
\left\{\begin{array}{l} 
n=0,1,2,3 \mbox{ for }I=0,2\,. \\
n=0,1,2 \mbox{ for }I=1\,.
\end{array}  \right.
\ee
These were the quantities used in Ref.~\cite{CGL1,CGL2} to perform
the matching of the Roy analysis with the $SU(2)$ ChPT constraints.
In addition they defined two combinations of these constants which had
very small nonanalytic contributions depending on the pion mass
when reexpressed with the help of the pion scalar radius.
These are given by
\ba
C_1 &=& F_\pi^2  \left(c_2+ 4 m_\pi^2\left(c_3-c_4\right)\right)\,,\nonumber\\
C_2 &=& \frac{F_\pi^2}{m_\pi^2}  
\left(-c_1+ 4 m_\pi^4\left(c_3-c_4\right)\right)\,.
\ea
They are both equal to one at lowest order. We have shown both $C_1$ and
$C_2$ as a function of $L_4^r$ and $L_6^r$ similarly to the plots shown
for $a^0_0$ and $a^2_0$ in Fig.~\ref{figC12}.
Also shown are the regions for them obtained
in Ref.~\cite{CGL2}.

\FIGURE[t]{
\begin{minipage}{0.48\textwidth}
\includegraphics[angle=270,width=0.99\textwidth]{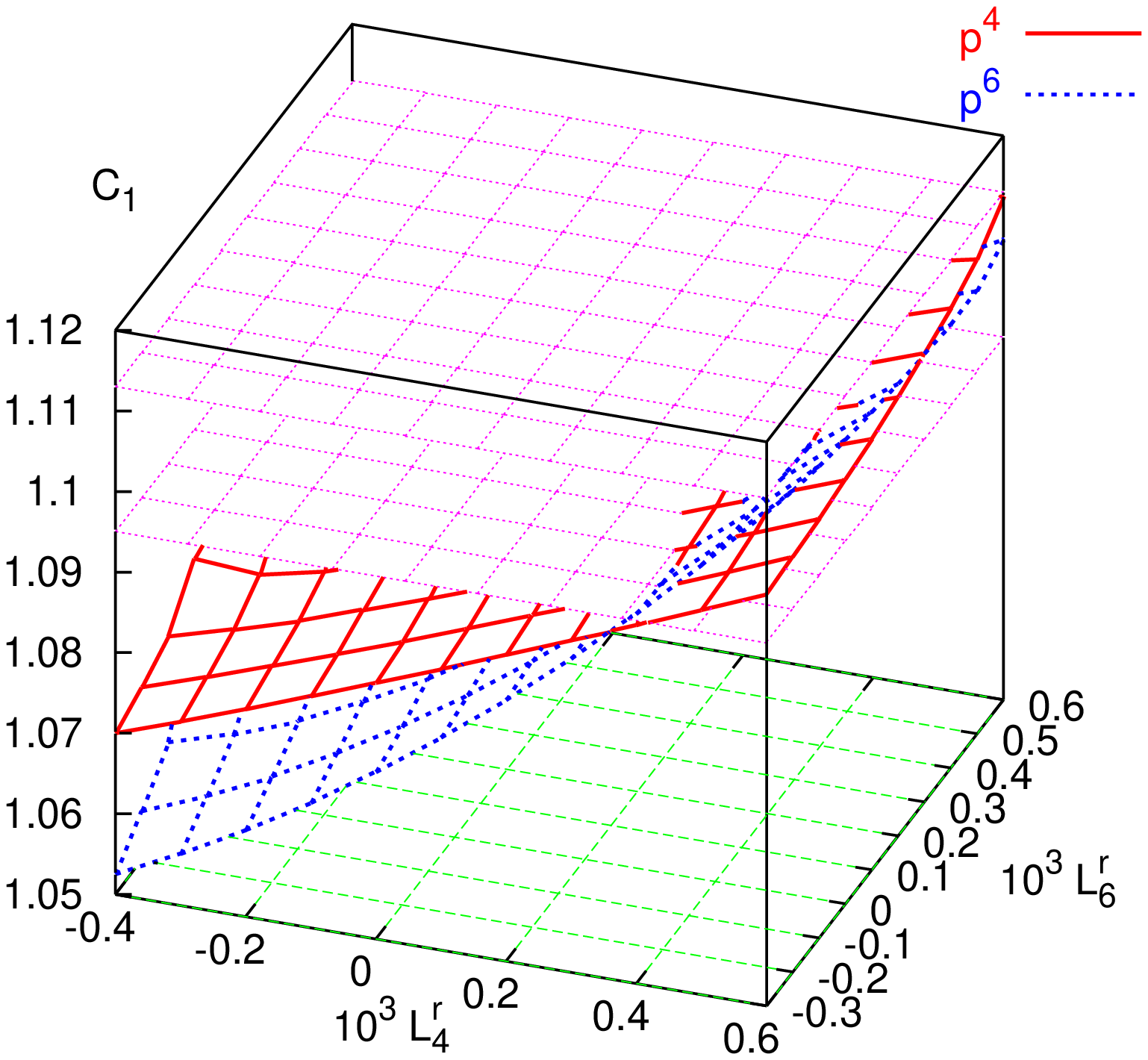}
\centerline{(a)}
\end{minipage}
\hfill
\begin{minipage}{0.48\textwidth}
\includegraphics[angle=270,width=0.99\textwidth]{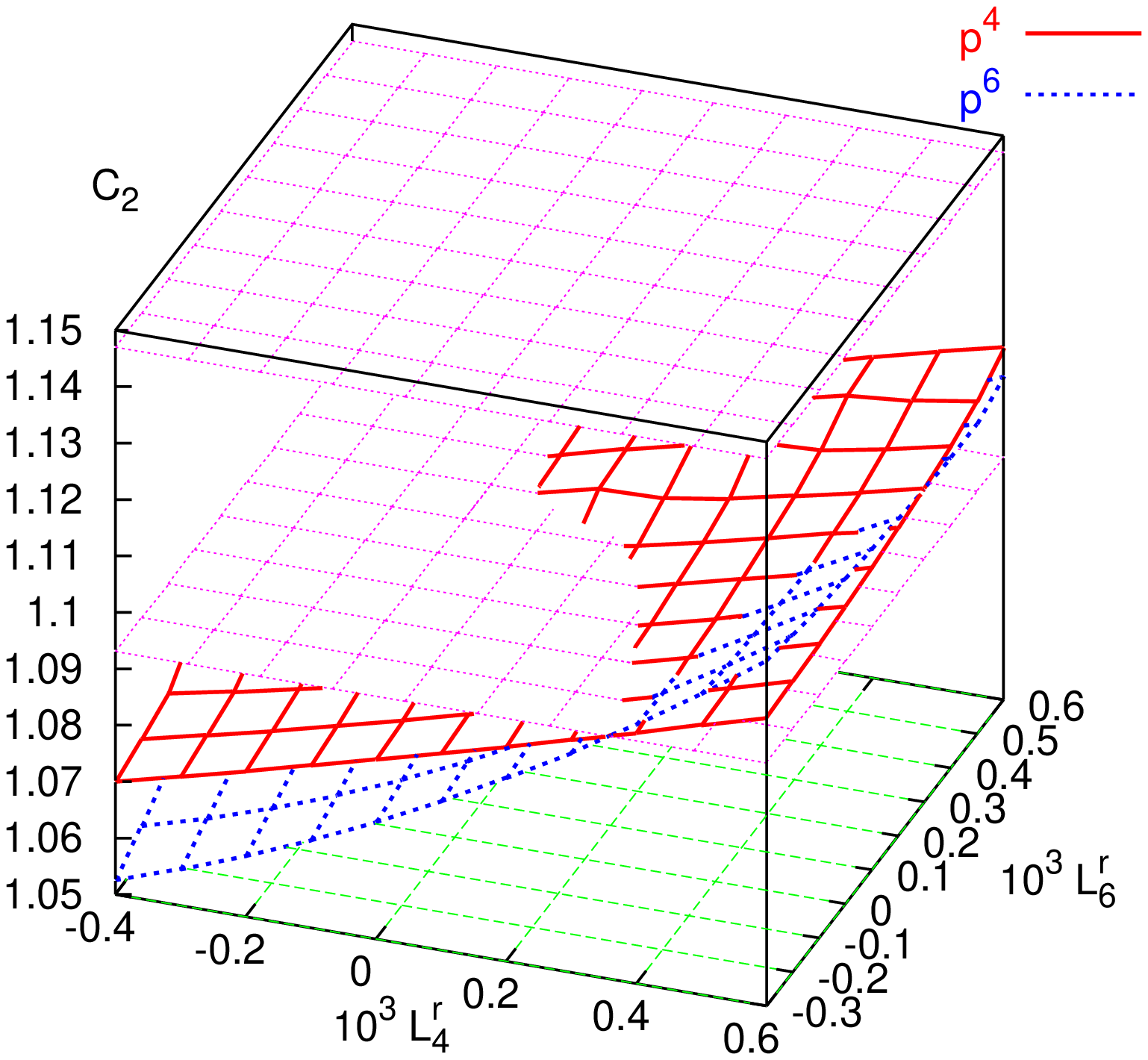}
\centerline{(b)}
\end{minipage}
\caption{\label{figC12} The subthreshold parameters
$C_1$ and $C_2$ as
a function of the input values $L_4^r$ and $L_6^r$ with the other $L_i^r$
simultaneously refitted to the $K_{e4}$ form factors.
(a) $C_1$ calculated to order $p^4$ and $p^6$.
(b) Idem for $C_2$. Notice that the lowest order value of 1 is outside the
plotted region.}
}

The plots show again that the preferred values of $L_4^r$ and $L_6^r$ are
somewhat different from zero in order to obtain good agreement.
The values of $c_5$ and $c_6$ can again be used to check the
resonance predictions for two possible combinations of the $C_i^r$.
These are in agreement to about a factor of two as expected.
The values for $c_3$  is in good agreement with the one obtained
by Ref.~\cite{CGL2} but the $p^6$ contribution is of the opposite
sign and larger than the $p^4$ contribution. For $c_4$ the agreement is
marginal,
similar to the one for $a^0_2$ shown in Table~\ref{tab:thresh2}.

\section{Conclusions}
\label{conclusions}

We have calculated $\pi\pi$ scattering to NNLO in three flavour ChPT
and presented the full expressions in App.~\ref{appB}. This is the main
results of this work. We then presented a first numerical analysis and
comparison with the low energy constants of order $p^4$ as determined
from the absolute values of the form factors in $K_{e4}$.

This comparison led to the conclusion that the preferred values of
$L_4^r$ and $L_6^r$ are somewhat different from zero but in size
compatible with the error estimates based on the arguments from the
large number of colours limit done in Ref.~\cite{GL2}. We have presented
plots of two of the threshold parameters as a function of $L_4^r$ and
$L_6^r$ with the other $L_i^r$ fitted to the $K_{e4}$ data.
The values for the threshold parameters have been presented for fit 10
of Ref.~\cite{ABT4} and fits A,B,C of Ref.~\cite{BD}. The latter are
those in the region preferred by the various scalar
form factor constraints. The convergence of the three flavour ChPT series
for $\pi\pi$ scattering seems reasonable and is similar to the one for
the two flavour case after the masses and decay constants have been put to
their physical values as was done here.

\acknowledgments

The program FORM 3.0 has been used extensively in these calculations
\cite{FORM}. This work is supported in part by the Swedish Research Council
and European Union TMR
network, Contract No. HPRN-CT-2002-00311  (EURIDICE).

\appendix
\section{Integrals}
\label{appA}
The one loop integrals we use are well known. The expressions for them can be
found in \cite{ABT1} and \cite{BT1}. The two loop integrals of sunset type
were derived in \cite{ABT1} using the results of \cite{PT} for the
subtraction constants. They are denoted by $\overline{H_i}$ in the expressions.

The two loop integrals of vertex type are evaluated using the methods
of \cite{Ghinculov}. Their precise definition and method of
evaluation can be found in \cite{BT1}. They are denoted by $\overline{V}$
in the expressions.

\section{Analytical results}
\label{appB}

We now write the amplitude in the form (\ref{defVi}) where we expand
all functions in the ChPT expansion which we label by a superscript
$(n)$ denoting the $p^n$ order in the expansion. The FORM files
with this output will be made available in \cite{formulas}.

The lowest order result is \cite{Weinbergpipi}
\be
V^{I(2)}(s) = 0\,,\quad
C(s,t,u)^{(2)} = -\frac{1}{F_\pi^2}\left(s-m_\pi^2\right)\,.
\ee
The order $p^4$ result agrees with the one shown in \cite{Knecht1}
and with the one in \cite{BKM1} up to a few misprints there.
The result is
\ba
32\pi F_\pi^4\,V^{0(4)}(s) &=&
       + \overline{B}(m_\pi^2,m_\pi^2,s) \, \left(
   - 2\,m_\pi^2\,s + 1/2\,m_\pi^4 + 2\,s^2 \right)
\nonumber\\&&
       + \overline{B}(m_K^2,m_K^2,s) \, \left( 3/8\,s^2 \right)
       + \overline{B}(m_\eta^2,m_\eta^2,s) \, \left( 1/6\,m_\pi^4 \right)
\,,\nonumber\\
32\pi F_\pi^4\,V^{1(4)}(s) &=&
       + \overline{B}(m_\pi^2,m_\pi^2,s) \, \left(  - 2/9\,m_\pi^2 + 1/18\,s \right)
\nonumber\\&&
       + \overline{B}(m_K^2,m_K^2,s) \, \left(  - 1/9\,m_K^2 + 1/36\,s \right)
\,,\nonumber\\
32\pi F_\pi^4\,V^{2(4)}(s) &=&
       + \overline{B}(m_\pi^2,m_\pi^2,s) \, \left(  - 2\,m_\pi^2\,s + 1/2\,m_\pi^4 + 2\,s^2 \right)
\nonumber\\&&
       + \overline{B}(m_K^2,m_K^2,s) \, \left( 3/8\,s^2 \right)
       + \overline{B}(m_\eta^2,m_\eta^2,s) \, \left( 1/6\,m_\pi^4 \right)
\,,
\ea     
and for the polynomial part
\ba
C(s,t,u)^{(4)} &=&
       + \frac{1}{16\pi^2} \, \left( 2/3\,m_\pi^2\,m_K^2 - 1/3\,m_\pi^2\,s + 2/3\,m_\pi^4 - 1/2\,m_K^2\,s - 1/
         8\,s^2 + 1/24\,\delta^2 \right)
\nonumber\\&&
       + \overline{A}(m_\pi^2) \, \left(  - 4/3\,m_\pi^2 + s \right)
       + \overline{A}(m_K^2) \, \left(  - 2/3\,m_\pi^2 + 1/2\,s \right)
\nonumber\\&&
       - 32\,m_\pi^2\,L^r_{1}s - 16\,m_\pi^2\,L^r_{3}s
  + 16\,m_\pi^2\,L^r_{4}s + 8\,m_\pi^2\,L^r_{5}s
\nonumber\\&&
          + 32\,m_\pi^4\,L^r_{1}+ 16\,m_\pi^4\,L^r_{3}
  - 32\,m_\pi^4\,L^r_{4}- 16\,m_\pi^4\,L^r_{5}
\nonumber\\&&
          + 32\,m_\pi^4\,L^r_{6}+ 16\,m_\pi^4\,L^r_{8}
  + 8\,L^r_{1}s^2 + 2\,L^r_{2}s^2
 + 2\,
         L^r_{2}\delta^2 + 4\,L^r_{3}s^2
\ea  
Here we used the notation
$\delta = t-u$ to have a shorter expression.

The order $p^6$ expressions are significantly longer. 
We split them in several parts
\ba
F_\pi^6\,C^{(6)}(s,t,u) &=& C_{\mathbf{C}}(s,t,u) + C_{\mathbf{L}}(s,t,u)
+C_{\mathbf{6}}(s,t,u)  \,,
\nonumber\\
32\pi F_\pi^6\, V^{I(6)}(s) &=&  V^I_{\mathbf{L}}(s)
+ V^I_{\mathbf{6}}(s)
+ V^I_{\mathbf{V}}(s)
\,.
\ea
The $C_i^r$ only
contribute to the polynomial part.
\ba
C_{\mathbf{C}}(s,t,u) &=&
       + s\,\delta^2 \, \left(
          + 6\,C^r_{3}
          + 2\,C^r_{4}
          \right)
\nonumber\\&&
       + s \, \Big(
          - 64\,m_\pi^2\,m_K^2\,C^r_{6}
          + 64\,m_\pi^2\,m_K^2\,C^r_{11}
          - 256\,m_\pi^2\,m_K^2\,C^r_{13}
          + 32\,m_\pi^2\,m_K^2\,C^r_{15}
\nonumber\\&&
          - 96\,m_\pi^4\,C^r_{1}
          - 192\,m_\pi^4\,C^r_{2}
          + 96\,m_\pi^4\,C^r_{3}
          + 32\,m_\pi^4\,C^r_{4}
          - 32\,m_\pi^4\,C^r_{5}
          - 32\,m_\pi^4\,C^r_{6}
\nonumber\\&&
          - 64\,m_\pi^4\,C^r_{7}
          - 32\,m_\pi^4\,C^r_{8}
          + 32\,m_\pi^4\,C^r_{10}
          + 32\,m_\pi^4\,C^r_{11}
          - 64\,m_\pi^4\,C^r_{12}
\nonumber\\&&
          + 32\,m_\pi^4\,C^r_{14}
          + 48\,m_\pi^4\,C^r_{15}
          + 64\,m_\pi^4\,C^r_{16}
          + 32\,m_\pi^4\,C^r_{17}
          - 32\,m_\pi^4\,C^r_{25}
\nonumber\\&&
          + 32\,m_\pi^4\,C^r_{26}
          + 64\,m_\pi^4\,C^r_{28}
          - 64\,m_\pi^4\,C^r_{29}
          - 32\,m_\pi^4\,C^r_{30}
          \Big)
\nonumber\\&&
       + s^2 \, \Big(
          + 48\,m_\pi^2\,C^r_{1}
          + 96\,m_\pi^2\,C^r_{2}
          - 48\,m_\pi^2\,C^r_{3}
          - 40\,m_\pi^2\,C^r_{4}
          + 8\,m_\pi^2\,C^r_{5}
\nonumber\\&&
          + 8\,m_\pi^2\,C^r_{6}
          + 16\,m_\pi^2\,C^r_{7}
          + 8\,m_\pi^2\,C^r_{8}
          - 4\,m_\pi^2\,C^r_{10}
          - 4\,m_\pi^2\,C^r_{11}
          + 8\,m_\pi^2\,C^r_{12}
\nonumber\\&&
          - 8\,m_\pi^2\,C^r_{13}
          - 8\,m_\pi^2\,C^r_{22}
          - 8\,m_\pi^2\,C^r_{23}
          + 20\,m_\pi^2\,C^r_{25}
\nonumber\\&&
          + 16\,m_K^2\,C^r_{6}
          - 8\,m_K^2\,C^r_{11}
          + 48\,m_K^2\,C^r_{13}
          \Big)
\nonumber\\&&
       + s^3 \, (
          - 8\,C^r_{1}
          - 16\,C^r_{2}
          + 10\,C^r_{3}
          + 14\,C^r_{4}
          )
\nonumber\\&&
       + \delta^2 \, \Big(
          - 8\,m_\pi^2\,C^r_{4}
          + 4\,m_\pi^2\,C^r_{10}
          + 4\,m_\pi^2\,C^r_{11}
          - 8\,m_\pi^2\,C^r_{12}
          - 8\,m_\pi^2\,C^r_{13}
\nonumber\\&&
          + 8\,m_\pi^2\,C^r_{22}
          + 8\,m_\pi^2\,C^r_{23}
          - 4\,m_\pi^2\,C^r_{25}
          + 8\,m_K^2\,C^r_{11}
          - 16\,m_K^2\,C^r_{13}
          \Big)
\nonumber\\&&
       + 64\,m_\pi^4\,m_K^2\,C^r_{6}
          - 64\,m_\pi^4\,m_K^2\,C^r_{11}
          + 192\,m_\pi^4\,m_K^2\,C^r_{13}
          - 64\,m_\pi^4\,m_K^2\,C^r_{15}
\nonumber\\&&
          + 64\,m_\pi^4\,m_K^2\,C^r_{20}
          + 384\,m_\pi^4\,m_K^2\,C^r_{21}
          + 64\,m_\pi^4\,m_K^2\,C^r_{32}
          + 64\,m_\pi^6\,C^r_{1}
\nonumber\\&&
          + 128\,m_\pi^6\,C^r_{2}
          - 64\,m_\pi^6\,C^r_{3}
          + 32\,m_\pi^6\,C^r_{5}
          + 32\,m_\pi^6\,C^r_{6}
          + 64\,m_\pi^6\,C^r_{7}
          + 32\,m_\pi^6\,C^r_{8}
\nonumber\\&&
          - 32\,m_\pi^6\,C^r_{10}
          - 32\,m_\pi^6\,C^r_{11}
          + 32\,m_\pi^6\,C^r_{12}
          - 32\,m_\pi^6\,C^r_{13}
          - 64\,m_\pi^6\,C^r_{14}
          - 96\,m_\pi^6\,C^r_{15}
\nonumber\\&&
          - 128\,m_\pi^6\,C^r_{16}
          - 64\,m_\pi^6\,C^r_{17}
          + 96\,m_\pi^6\,C^r_{19}
          + 160\,m_\pi^6\,C^r_{20}
          + 192\,m_\pi^6\,C^r_{21}
\nonumber\\&&
          - 64\,m_\pi^6\,C^r_{26}
          - 128\,m_\pi^6\,C^r_{28}
          + 64\,m_\pi^6\,C^r_{29}
          + 96\,m_\pi^6\,C^r_{31}
          + 160\,m_\pi^6\,C^r_{32}\,.
\ea

\ba
C_{\mathbf{L}}(s,t,u) &=&
       + \frac{1}{16\pi^2} \, \Big(
          + 256/27\,m_\pi^2\,m_K^2\,L^r_{2}s
          + 1048/81\,m_\pi^2\,m_K^2\,L^r_{3}s
          - 8\,m_\pi^2\,m_K^2\,L^r_{5}s
\nonumber\\&&
          - 92/9\,m_\pi^2\,L^r_{1}s^2
          + 28/9\,m_\pi^2\,L^r_{1}\delta^2
          - 386/27\,m_\pi^2\,L^r_{2}s^2
          - 2/9\,m_\pi^2\,L^r_{2}\delta^2
\nonumber\\&&
          - 320/81\,m_\pi^2\,L^r_{3}s^2
          + 4/3\,m_\pi^2\,L^r_{3}\delta^2
          - 8/3\,m_\pi^2\,L^r_{4}s^2
          + 8/9\,m_\pi^2\,L^r_{4}\delta^2
          - 2\,m_\pi^2\,L^r_{5}s^2
\nonumber\\&&
          + 2/3\,m_\pi^2\,L^r_{5}\delta^2
          + 128/27\,m_\pi^4\,m_K^2\,L^r_{2}
          - 736/81\,m_\pi^4\,m_K^2\,L^r_{3}
          + 32/3\,m_\pi^4\,m_K^2\,L^r_{5}
\nonumber\\&&
          + 176/3\,m_\pi^4\,L^r_{1}s
          + 808/27\,m_\pi^4\,L^r_{2}s
          + 2008/81\,m_\pi^4\,L^r_{3}s
          - 160/9\,m_\pi^4\,L^r_{4}s
\nonumber\\&&
          - 16/3\,m_\pi^4\,L^r_{5}s
          - 448/9\,m_\pi^6\,L^r_{1}
          - 224/27\,m_\pi^6\,L^r_{2}
          - 1760/81\,m_\pi^6\,L^r_{3}
\nonumber\\&&
          + 256/9\,m_\pi^6\,L^r_{4}
          + 32/3\,m_\pi^6\,L^r_{5}
          - 8/3\,m_K^2\,L^r_{2}s^2
          - 8/3\,m_K^2\,L^r_{3}s^2
          + 2/3\,m_K^2\,L^r_{3}\delta^2
\nonumber\\&&
          + 1/9\,L^r_{1}s\,\delta^2
          + 1/3\,L^r_{1}s^3
          + 5/18\,L^r_{2}s\,\delta^2
          + 13/6\,L^r_{2}s^3
          + 1/9\,L^r_{3}s^3
          \Big)
\nonumber\\&&
       + \overline{A}(m_\eta^2) \, \Big(
          - 16/3\,m_\pi^2\,L^r_{1}s
          - 16/9\,m_\pi^2\,L^r_{2}s
          - 40/27\,m_\pi^2\,L^r_{3}s
          + 32/3\,m_\pi^4\,L^r_{1}
\nonumber\\&&
          + 16/9\,m_\pi^4\,L^r_{2}
          + 64/27\,m_\pi^4\,L^r_{3}
          - 16/3\,m_\pi^4\,L^r_{4}
          - 8/9\,m_\pi^4\,L^r_{5}
          + 32/3\,m_\pi^4\,L^r_{7}
\nonumber\\&&
          + 16/3\,m_\pi^4\,L^r_{8}
          \Big)
\nonumber\\&&
       + \overline{A}(m_\pi^2) \, \Big(
          - 32/3\,m_\pi^2\,L^r_{1}s
          + 184/3\,m_\pi^2\,L^r_{2}s
          - 16/3\,m_\pi^2\,L^r_{3}s
          + 64\,m_\pi^2\,L^r_{4}s
\nonumber\\&&
          + 32\,m_\pi^2\,L^r_{5}s
          + 64\,m_\pi^4\,L^r_{1}
          - 32\,m_\pi^4\,L^r_{2}
          + 32\,m_\pi^4\,L^r_{3}
          - 464/3\,m_\pi^4\,L^r_{4}
\nonumber\\&&
          - 232/3\,m_\pi^4\,L^r_{5}
          + 160\,m_\pi^4\,L^r_{6}
          + 80\,m_\pi^4\,L^r_{8}
          - 12\,L^r_{1}s^2
          + 4\,L^r_{1}\delta^2
\nonumber\\&&
          - 50/3\,L^r_{2}s^2
          + 38/3\,L^r_{2}\delta^2
          - 6\,L^r_{3}s^2
          + 2\,L^r_{3}\delta^2
          \Big)
\nonumber\\&&
       + \overline{A}(m_K^2) \, \Big(
          - 32\,m_\pi^2\,L^r_{1}s
          + 16/3\,m_\pi^2\,L^r_{2}s
          - 52/3\,m_\pi^2\,L^r_{3}s
          + 16\,m_\pi^2\,L^r_{4}s
\nonumber\\&&
          + 16\,m_\pi^2\,L^r_{5}s
          + 64\,m_\pi^4\,L^r_{1}
          + 80/3\,m_\pi^4\,L^r_{3}
          - 64\,m_\pi^4\,L^r_{4}
          - 104/3\,m_\pi^4\,L^r_{5}
\nonumber\\&&
          + 64\,m_\pi^4\,L^r_{6}
          + 32\,m_\pi^4\,L^r_{8}
          - 4/3\,L^r_{2}s^2
          + 4\,L^r_{2}\delta^2
          + 5/3\,L^r_{3}s^2
          + 1/3\,L^r_{3}\delta^2
          \Big)
\nonumber\\&&
       + 128\,m_\pi^2\,m_K^2\,L^r_{4}L^r_{5}s
          - 512\,m_\pi^2\,m_K^2\,L^r_{4}L^r_{6}s
          + 256\,m_\pi^2\,m_K^2\,(L^r_{4})^2\,s
\nonumber\\&&
          - 256\,m_\pi^2\,m_K^2\,L^r_{5}L^r_{6}s
          - 256\,m_\pi^4\,m_K^2\,L^r_{4}L^r_{5}
          + 2048\,m_\pi^4\,m_K^2\,L^r_{4}L^r_{6}
\nonumber\\&&
          + 512\,m_\pi^4\,m_K^2\,L^r_{4}L^r_{8}
          - 512\,m_\pi^4\,m_K^2\,(L^r_{4})^2
          + 512\,m_\pi^4\,m_K^2\,L^r_{5}L^r_{6}
\nonumber\\&&
          - 1024\,m_\pi^4\,m_K^2\,L^r_{6}L^r_{8}
          - 2048\,m_\pi^4\,m_K^2\,(L^r_{6})^2
          + 192\,m_\pi^4\,L^r_{4}L^r_{5}s
\nonumber\\&&
          - 256\,m_\pi^4\,L^r_{4}L^r_{6}s
          - 256\,m_\pi^4\,L^r_{4}L^r_{8}s
          + 128\,m_\pi^4\,(L^r_{4})^2\,s
          - 128\,m_\pi^4\,L^r_{5}L^r_{6}s
\nonumber\\&&
          - 128\,m_\pi^4\,L^r_{5}L^r_{8}s
          + 64\,m_\pi^4\,(L^r_{5})^2\,s
          - 384\,m_\pi^6\,L^r_{4}L^r_{5}
          + 1024\,m_\pi^6\,L^r_{4}L^r_{6}
\nonumber\\&&
          + 768\,m_\pi^6\,L^r_{4}L^r_{8}
          - 256\,m_\pi^6\,(L^r_{4})^2
          + 768\,m_\pi^6\,L^r_{5}L^r_{6}
          + 512\,m_\pi^6\,L^r_{5}L^r_{8}
\nonumber\\&&
          - 128\,m_\pi^6\,(L^r_{5})^2
          - 1536\,m_\pi^6\,L^r_{6}L^r_{8}
          - 1024\,m_\pi^6\,(L^r_{6})^2
          - 512\,m_\pi^6\,(L^r_{8})^2\,.
\ea

\ba
 C_{\mathbf{6}}(s,t,u) &=&
       + \frac{1}{16\pi^2}\,\overline{A}(m_\eta^2) \, \Big(
          + 1/36\,m_\pi^2\,s
          - 1/9\,m_\pi^4
          \Big)
\nonumber\\&&
       + \frac{1}{16\pi^2}\,\overline{A}(m_\pi^2) \, \Big(
          + 8/9\,m_\pi^2\,m_K^2
          - 89/36\,m_\pi^2\,s
          + 29/18\,m_\pi^4
          - 2/3\,m_K^2\,s
\nonumber\\&&
          + 5/24\,s^2
          - 1/72\,\delta^2
          \Big)
\nonumber\\&&
       + \frac{1}{16\pi^2}\,\overline{A}(m_K^2) \, \Big(
          + 4/9\,m_\pi^2\,m_K^2
          - 37/36\,m_\pi^2\,s
          + 3/4\,m_\pi^4
          - 1/3\,m_K^2\,s
\nonumber\\&&
          + 1/24\,s^2
          + 1/72\,\delta^2
          \Big)
\nonumber\\&&
       + \left(\frac{1}{16\pi^2}\right)^2 \, \Big(
          - 83/216\,m_\pi^2\,m_K^2\,pi^2\,s
          - 1375/432\,m_\pi^2\,m_K^2\,s
          - 2/9\,m_\pi^2\,m_K^4
\nonumber\\&&
          + 13/32\,m_\pi^2\,\pi^2\,s^2
          - 19/288\,m_\pi^2\,pi^2\,\delta^2
          + 1751/432\,m_\pi^2\,s^2
          - 289/432\,m_\pi^2\,\delta^2
\nonumber\\&&
          + 103/648\,m_\pi^4\,m_K^2\,\pi^2
          + 167/432\,m_\pi^4\,m_K^2
          - 337/216\,m_\pi^4\,\pi^2\,s
\nonumber\\&&
          - 10933/864\,m_\pi^4\,s
          + 349/324\,m_\pi^6\,\pi^2
          + 6013/864\,m_\pi^6
          + 1/8\,m_K^2\,\pi^2\,s^2
\nonumber\\&&
          - 1/144\,m_K^2\,\pi^2\,\delta^2
          + 13/12\,m_K^2\,s^2
          - 7/72\,m_K^2\,\delta^2
          + 1/6\,m_K^4\,s
\nonumber\\&&
          - 1/96\,s\,\delta^2
          - 11/96\,s^3
          \Big)
\nonumber\\&&
       + \overline{A}(m_\eta^2)^2 \, \Big(
          + 1/12\,m_\pi^2\,m_\eta^{-2}\,s
          - 2/9\,m_\pi^4\,m_\eta^{-2}
          \Big)
\nonumber\\&&
       + \overline{A}(m_\pi^2)\,\overline{A}(m_K^2) \, \Big(
          - 8/9\,m_\pi^2
          + 2/3\,s
          \Big)
\nonumber\\&&
       + \overline{A}(m_\pi^2)^2 \, \Big(
          + 39/16\,m_\pi^{-2}\,s^2
          - 19/48\,m_\pi^{-2}\,\delta^2
          + 11/2\,m_\pi^2
          - 26/3\,s
          \Big)
\nonumber\\&&
       + \overline{A}(m_K^2)^2 \, \Big(
          - 29/12\,m_\pi^2\,m_K^{-2}\,s
          - 2/9\,m_\pi^2
          + 5/4\,m_\pi^4\,m_K^{-2}
          + 3/4\,m_K^{-2}\,s^2
\nonumber\\&&
          - 1/24\,m_K^{-2}\,\delta^2
          + 1/6\,s
          \Big)
\nonumber\\&&
       + H^{F\prime}(m_\pi^2,m_\pi^2,m_\pi^2,m_\pi^2) \, \Big(
          + 5/2\,m_\pi^4\,s
          - 5/2\,m_\pi^6
          \Big)
\nonumber\\&&
       + H^{F\prime}(m_\pi^2,m_K^2,m_K^2,m_\pi^2) \, \Big(
          - 15/8\,m_\pi^4\,s
          + 15/8\,m_\pi^6
          \Big)
\nonumber\\&&
       + H^{F\prime}(m_\pi^2,m_\eta^2,m_\eta^2,m_\pi^2) \, \Big(
          + 1/6\,m_\pi^4\,s
          - 1/6\,m_\pi^6
          \Big)
\nonumber\\&&
       + H^{F\prime}(m_K^2,m_\pi^2,m_K^2,m_\pi^2) \, \Big(
          + 3\,m_\pi^2\,m_K^2\,s
          - 3\,m_\pi^4\,m_K^2
          \Big)
\nonumber\\&&
       + H^{F\prime}(m_K^2,m_K^2,m_\eta^2,m_\pi^2) \, \Big(
          - 5/2\,m_\pi^4\,s
          + 5/2\,m_\pi^6
          \Big)
\nonumber\\&&
       + H^{F\prime}(m_\eta^2,m_K^2,m_K^2,m_\pi^2) \, \Big(
          + 3/2\,m_\pi^2\,m_K^2\,s
          - 3/2\,m_\pi^4\,m_K^2
          + 13/8\,m_\pi^4\,s
\nonumber\\&&
          - 13/8\,m_\pi^6
          \Big)
\nonumber\\&&
       + H^{F\prime}_1(m_\pi^2,m_K^2,m_K^2,m_\pi^2) \, \Big(
          + 3\,m_\pi^4\,s
          - 3\,m_\pi^6
          \Big)
\nonumber\\&&
       + H^{F\prime}_1(m_K^2,m_K^2,m_\eta^2,m_\pi^2) \, \Big(
          + 2\,m_\pi^4\,s
          - 2\,m_\pi^6
          \Big)
\nonumber\\&&
       + H^{F\prime}_1(m_\eta^2,m_K^2,m_K^2,m_\pi^2) \, \Big(
          - 2\,m_\pi^4\,s
          + 2\,m_\pi^6
          \Big)
\nonumber\\&&
       + H^{F\prime}_{21}(m_\pi^2,m_\pi^2,m_\pi^2,m_\pi^2) \, \Big(
          + 9\,m_\pi^4\,s
          - 9\,m_\pi^6
          \Big)
\nonumber\\&&
       + H^{F\prime}_{21}(m_\pi^2,m_K^2,m_K^2,m_\pi^2) \, \Big(
          - 9/8\,m_\pi^4\,s
          + 9/8\,m_\pi^6
          \Big)
\nonumber\\&&
       + H^{F\prime}_{21}(m_K^2,m_\pi^2,m_K^2,m_\pi^2) \, \Big(
          + 9\,m_\pi^4\,s
          - 9\,m_\pi^6
          \Big)
\nonumber\\&&
       + H^{F\prime}_{21}(m_\eta^2,m_K^2,m_K^2,m_\pi^2) \, \Big(
          + 27/8\,m_\pi^4\,s
          - 27/8\,m_\pi^6
          \Big)
\nonumber\\&&
       + H^{F}(m_\pi^2,m_\pi^2,m_\pi^2,m_\pi^2) \, \Big(
          + 4\,m_\pi^2\,s
          - 17/6\,m_\pi^4
          - s^2
          + 1/3\,\delta^2
          \Big)
\nonumber\\&&
       + H^{F}(m_\pi^2,m_K^2,m_K^2,m_\pi^2) \, \Big(
          - 133/36\,m_\pi^2\,s
          + 10/3\,m_\pi^4
          - 5/16\,s^2
          + 7/16\,\delta^2
          \Big)
\nonumber\\&&
       + H^{F}(m_\pi^2,m_\eta^2,m_\eta^2,m_\pi^2) \, \Big(
          + 1/18\,m_\pi^4
          \Big)
\nonumber\\&&
       + H^{F}(m_K^2,m_\pi^2,m_K^2,m_\pi^2) \, \Big(
          - 1/12\,m_\pi^2\,m_K^2
          + 53/18\,m_\pi^2\,s
          - 4/3\,m_\pi^4
\nonumber\\&&
          + 13/72\,m_K^2\,s
          - 1/4\,s^2
          - 1/4\,\delta^2
          \Big)
\nonumber\\&&
       + H^{F}(m_K^2,m_K^2,m_\pi^2,m_\pi^2) \, \Big(
          + 1/12\,m_\pi^2\,m_K^2
          - 13/72\,m_K^2\,s
          \Big)
\nonumber\\&&
       + H^{F}(m_K^2,m_K^2,m_\eta^2,m_\pi^2) \, \Big(
          - 7/18\,m_\pi^4
          + 3/16\,s^2
          - 3/16\,\delta^2
          \Big)
\nonumber\\&&
       + H^{F}(m_\eta^2,m_K^2,m_K^2,m_\pi^2) \, \Big(
          - 11/8\,m_\pi^2\,s
          + 119/36\,m_\pi^4
          - 45/64\,s^2
          + 27/64\,\delta^2
          \Big)
\nonumber\\&&
       + H^{F}_1(m_\pi^2,m_K^2,m_K^2,m_\pi^2) \, \Big(
          + 7/4\,m_\pi^2\,s
          - 3/2\,m_\pi^4
          - 3/16\,s^2
          + 1/16\,\delta^2
          \Big)
\nonumber\\&&
       + H^{F}_1(m_\eta^2,m_K^2,m_K^2,m_\pi^2) \, \Big(
          + 11/8\,m_\pi^2\,s
          - 15/4\,m_\pi^4
          + 45/64\,s^2
          - 27/64\,\delta^2
          \Big)
\nonumber\\&&
       + H^{F}_{21}(m_\pi^2,m_\pi^2,m_\pi^2,m_\pi^2) \, \Big(
          + 5\,m_\pi^2\,s
          - m_\pi^4
          - 3/2\,s^2
          + 1/2\,\delta^2
          \Big)
\nonumber\\&&
       + H^{F}_{21}(m_\pi^2,m_K^2,m_K^2,m_\pi^2) \, \Big(
          - 13/8\,m_\pi^2\,s
          + 1/8\,m_\pi^4
          + 15/16\,s^2
          - 5/16\,\delta^2
          \Big)
\nonumber\\&&
       + H^{F}_{21}(m_K^2,m_\pi^2,m_K^2,m_\pi^2) \, \Big(
          + 253/24\,m_\pi^2\,s
          - 45/4\,m_\pi^4
          + 3/8\,s^2
          - 1/8\,\delta^2
          \Big)
\nonumber\\&&
       + H^{F}_{21}(m_K^2,m_K^2,m_\pi^2,m_\pi^2) \, \Big(
          - 37/24\,m_\pi^2\,s
          + 17/4\,m_\pi^4
          - 3/2\,s^2
          + 1/2\,\delta^2
          \Big)
\nonumber\\&&
       + H^{F}_{21}(m_K^2,m_K^2,m_\eta^2,m_\pi^2) \, \Big(
          + 3/2\,m_\pi^2\,s
          - 9/2\,m_\pi^4
          + 45/32\,s^2
          - 15/32\,\delta^2
          \Big)
\nonumber\\&&
       + H^{F}_{21}(m_\eta^2,m_K^2,m_K^2,m_\pi^2) \, \Big(
          + 21/8\,m_\pi^2\,s
          - 3/8\,m_\pi^4
          - 9/8\,s^2
          + 3/8\,\delta^2
          \Big)\,.
\ea

\ba
V^0_{\mathbf{L}}(s) &=&
       + \overline{B}(m_\pi^2,m_\pi^2,s) \, \Big(
          - 728/3\,m_\pi^2\,L^r_{1}s^2
          - 376/3\,m_\pi^2\,L^r_{2}s^2
          - 364/3\,m_\pi^2\,L^r_{3}s^2
\nonumber\\&&
          + 64\,m_\pi^2\,L^r_{4}s^2
          + 32\,m_\pi^2\,L^r_{5}s^2
          + 1024/3\,m_\pi^4\,L^r_{1}s
          + 608/3\,m_\pi^4\,L^r_{2}s
          + 512/3\,m_\pi^4\,L^r_{3}s
\nonumber\\&&
          - 224\,m_\pi^4\,L^r_{4}s
          - 112\,m_\pi^4\,L^r_{5}s
          + 320\,m_\pi^4\,L^r_{6}s
          + 160\,m_\pi^4\,L^r_{8}s
          - 352/3\,m_\pi^6\,L^r_{1}
\nonumber\\&&
          - 224/3\,m_\pi^6\,L^r_{2}
          - 176/3\,m_\pi^6\,L^r_{3}
          + 96\,m_\pi^6\,L^r_{4}
          + 48\,m_\pi^6\,L^r_{5}
          - 160\,m_\pi^6\,L^r_{6}
\nonumber\\&&
          - 80\,m_\pi^6\,L^r_{8}
          + 176/3\,L^r_{1}s^3
          + 112/3\,L^r_{2}s^3
          + 88/3\,L^r_{3}s^3
          \Big)
\nonumber\\&&
       + \overline{B}(m_K^2,m_K^2,s) \, \Big(
          + 96\,m_\pi^2\,m_K^2\,L^r_{1}s
          + 32\,m_\pi^2\,m_K^2\,L^r_{2}s
          + 32\,m_\pi^2\,m_K^2\,L^r_{3}s
\nonumber\\&&
          - 96\,m_\pi^2\,m_K^2\,L^r_{4}s
          - 24\,m_\pi^2\,m_K^2\,L^r_{5}s
          + 96\,m_\pi^2\,m_K^2\,L^r_{6}s
          + 48\,m_\pi^2\,m_K^2\,L^r_{8}s
\nonumber\\&&
          - 48\,m_\pi^2\,L^r_{1}s^2
          - 8\,m_\pi^2\,L^r_{2}s^2
          - 14\,m_\pi^2\,L^r_{3}s^2
          + 24\,m_\pi^2\,L^r_{4}s^2
          + 6\,m_\pi^2\,L^r_{5}s^2
\nonumber\\&&
          - 48\,m_K^2\,L^r_{1}s^2
          - 8\,m_K^2\,L^r_{2}s^2
          - 14\,m_K^2\,L^r_{3}s^2
          + 24\,m_K^2\,L^r_{4}s^2
          + 24\,L^r_{1}s^3
\nonumber\\&&
          + 8\,L^r_{2}s^3
          + 8\,L^r_{3}s^3
          \Big)
\nonumber\\&&
       + \overline{B}(m_\eta^2,m_\eta^2,s) \, \Big(
          - 64/3\,m_\pi^2\,m_K^2\,L^r_{1}s
          - 32/9\,m_\pi^2\,m_K^2\,L^r_{2}s
          - 128/27\,m_\pi^2\,m_K^2\,L^r_{3}s
\nonumber\\&&
          + 32/3\,m_\pi^2\,m_K^2\,L^r_{4}s
          + 8\,m_\pi^2\,L^r_{1}s^2
          + 8/3\,m_\pi^2\,L^r_{2}s^2
          + 20/9\,m_\pi^2\,L^r_{3}s^2
\nonumber\\&&
          + 128/3\,m_\pi^4\,m_K^2\,L^r_{1}
          + 128/9\,m_\pi^4\,m_K^2\,L^r_{2}
          + 320/27\,m_\pi^4\,m_K^2\,L^r_{3}
          - 128/3\,m_\pi^4\,m_K^2\,L^r_{4}
\nonumber\\&&
          - 64/9\,m_\pi^4\,m_K^2\,L^r_{5}
          + 128/3\,m_\pi^4\,m_K^2\,L^r_{6}
          - 128/3\,m_\pi^4\,m_K^2\,L^r_{7}
          - 32/3\,m_\pi^4\,L^r_{1}s
\nonumber\\&&
          - 16/9\,m_\pi^4\,L^r_{2}s
          - 64/27\,m_\pi^4\,L^r_{3}s
          + 16/3\,m_\pi^4\,L^r_{4}s
          - 32/3\,m_\pi^6\,L^r_{1}
          - 32/9\,m_\pi^6\,L^r_{2}
\nonumber\\&&
          - 80/27\,m_\pi^6\,L^r_{3}
          + 32/3\,m_\pi^6\,L^r_{4}
          + 16/9\,m_\pi^6\,L^r_{5}
          - 32/3\,m_\pi^6\,L^r_{6}
          + 128/3\,m_\pi^6\,L^r_{7}
\nonumber\\&&
          + 16\,m_\pi^6\,L^r_{8}
          \Big)\,.
\ea

\ba
V^0_{\mathbf{6}}(s) &=&
       + \overline{B}(m_\pi^2,m_\pi^2,s)\,\frac{1}{16\pi^2} \, \Big(
          + 85/24\,m_\pi^2\,s^2
          - 25/3\,m_\pi^4\,s
          + 10/3\,m_\pi^6
          - 5/12\,s^3
          \Big)
\nonumber\\&&
       + \overline{B}(m_\pi^2,m_\pi^2,s)^2 \, \Big(
          - 3\,m_\pi^2\,s^2
          + 3/2\,m_\pi^4\,s
          - 1/4\,m_\pi^6
          + 2\,s^3
          \Big)
\nonumber\\&&
       + \overline{B}(m_\pi^2,m_\pi^2,s)\,\overline{B}(m_K^2,m_K^2,s) \, \Big(
          - 3/8\,m_\pi^2\,s^2
          + 3/4\,s^3
          \Big)
\nonumber\\&&
       + \overline{B}(m_\pi^2,m_\pi^2,s)\,\overline{B}(m_\eta^2,m_\eta^2,s)
 \, \Big(
          + 1/3\,m_\pi^4\,s
          - 1/6\,m_\pi^6
          \Big)
\nonumber\\&&
       + \overline{B}(m_\pi^2,m_\pi^2,s) \, \Big(
          - 20\,m_\pi^2\,s\,\overline{A}(m_\pi^2)
          - 10\,m_\pi^2\,s\,\overline{A}(m_K^2)
          + 8\,m_\pi^4\,\overline{A}(m_\pi^2)
\nonumber\\&&
          + 4\,m_\pi^4\,\overline{A}(m_K^2)
          + 8\,s^2\,\overline{A}(m_\pi^2)
          + 4\,s^2\,\overline{A}(m_K^2)
          \Big)
\nonumber\\&&
       + \overline{B}(m_K^2,m_K^2,s)\,\frac{1}{16\pi^2} \, \Big(
          - 2\,m_\pi^2\,m_K^2\,s
          + 1/2\,m_\pi^2\,s^2
          + 1/2\,m_K^2\,s^2
          - 1/8\,s^3
          \Big)
\nonumber\\&&
       + \overline{B}(m_K^2,m_K^2,s)^2 \, \Big(
          + 9/32\,s^3
          \Big)
\nonumber\\&&
       + \overline{B}(m_K^2,m_K^2,s)\,\overline{B}(m_\eta^2,m_\eta^2,s)
 \, \Big(\nonumber\\&&
          - 1/3\,m_\pi^2\,m_K^2\,s
          + 3/8\,m_\pi^2\,s^2
          \Big)
\nonumber\\&&
       + \overline{B}(m_K^2,m_K^2,s) \, \Big(
          - 3/4\,m_\pi^2\,\overline{A}(m_\eta^2)\,s
          - 9/4\,m_\pi^2\,s\,\overline{A}(m_\pi^2)
          - 3/2\,m_\pi^2\,s\,\overline{A}(m_K^2)
\nonumber\\&&
          + 3/2\,s^2\,\overline{A}(m_\pi^2)
          + 3/4\,s^2\,\overline{A}(m_K^2)
          \Big)
\nonumber\\&&
       + \overline{B}(m_\eta^2,m_\eta^2,s)\,\frac{1}{16\pi^2} \, \Big(
          + 2/9\,m_\pi^2\,m_K^2\,s
          - 1/24\,m_\pi^2\,s^2
          - 8/9\,m_\pi^4\,m_K^2
          + 1/9\,m_\pi^4\,s
\nonumber\\&&
          + 2/9\,m_\pi^6
          \Big)
\nonumber\\&&
       + \overline{B}(m_\eta^2,m_\eta^2,s)^2 \, \Big(
          + 4/27\,m_\pi^4\,m_K^2
          - 7/108\,m_\pi^6
          \Big)
\nonumber\\&&
       + \overline{B}(m_\eta^2,m_\eta^2,s) \, \Big(
          + 2/3\,m_\pi^4\,\overline{A}(m_\pi^2)
          - 2/3\,m_\pi^4\,\overline{A}(m_K^2)
          \Big)
\nonumber\\&&
       + \overline{B}^\epsilon(m_\pi^2,m_\pi^2,s)\,\frac{1}{16\pi^2} \, \Big(
          - 10\,m_\pi^2\,m_K^2\,s
          + 59/4\,m_\pi^2\,s^2
          + 4\,m_\pi^4\,m_K^2
\nonumber\\&&
          - 335/9\,m_\pi^4\,s
          + 136/9\,m_\pi^6
          + 4\,m_K^2\,s^2
          - 3/2\,s^3
          \Big)
\nonumber\\&&
       + \overline{B}^\epsilon(m_K^2,m_K^2,s)\,\frac{1}{16\pi^2} \, \Big(
          - 6\,m_\pi^2\,m_K^2\,s
          + 15/8\,m_\pi^2\,s^2
          - 2\,m_\pi^4\,s
          + 9/4\,m_K^2\,s^2
\nonumber\\&&
          - 9/16\,s^3
          \Big)
\nonumber\\&&
       + \overline{B}^\epsilon(m_\eta^2,m_\eta^2,s)\,\frac{1}{16\pi^2} \, \Big(
          + 1/3\,m_\pi^2\,m_K^2\,s
          - 1/4\,m_\pi^2\,s^2
          - 4/3\,m_\pi^4\,m_K^2
          + 2/3\,m_\pi^4\,s
\nonumber\\&&
          - 2/9\,m_\pi^6
          \Big)\,.
\ea

\ba
V^0_{\mathbf{V}}(s)&=&
       + V_{0}(m_\pi^2,m_\pi^2,m_\pi^2,m_\pi^2,m_\pi^2,s,m_\pi^2) \, \Big(
          + 4\,m_\pi^2\,s^2
          - 4\,m_\pi^4\,s
          + m_\pi^6
          \Big)
\nonumber\\&&
       + V_{0}(m_\pi^2,m_\pi^2,m_K^2,m_K^2,m_\pi^2,s,m_\pi^2) \, \Big(
          + 1/2\,m_\pi^2\,s^2
          - 9/4\,m_\pi^4\,s
          + m_\pi^6
          \Big)
\nonumber\\&&
       + V_{0}(m_\pi^2,m_\pi^2,m_\eta^2,m_\eta^2,m_\pi^2,s,m_\pi^2) \, \Big(
          + 2/9\,m_\pi^4\,s
          - 1/9\,m_\pi^6
          \Big)
\nonumber\\&&
       + V_{0}(m_K^2,m_K^2,m_\pi^2,m_K^2,m_\pi^2,s,m_\pi^2) \, \Big(
          + 9/2\,m_\pi^4\,s
          \Big)
\nonumber\\&&
       + V_{0}(m_K^2,m_K^2,m_K^2,m_\eta^2,m_\pi^2,s,m_\pi^2) \, \Big(
          - 3/8\,m_\pi^2\,s^2
          + 2\,m_\pi^4\,s
          \Big)
\nonumber\\&&
       + V_{0}(m_\eta^2,m_\eta^2,m_\pi^2,m_\eta^2,m_\pi^2,s,m_\pi^2) \, \Big(
          + 2/9\,m_\pi^6
          \Big)
\nonumber\\&&
       + V_{0}(m_\eta^2,m_\eta^2,m_K^2,m_K^2,m_\pi^2,s,m_\pi^2) \, \Big(
          + 1/4\,m_\pi^4\,s
          + 1/3\,m_\pi^6
          \Big)
\nonumber\\&&
       + V_{11}(m_\pi^2,m_\pi^2,m_K^2,m_K^2,m_\pi^2,s,m_\pi^2) \, \Big(
          - m_\pi^2\,s^2
          + 9/2\,m_\pi^4\,s
          - 2\,m_\pi^6
          \Big)
\nonumber\\&&
       + V_{11}(m_K^2,m_K^2,m_\pi^2,m_K^2,m_\pi^2,s,m_\pi^2) \, \Big(
          + 3\,m_\pi^2\,s^2
          - 12\,m_\pi^4\,s
          \Big)
\nonumber\\&&
       + V_{11}(m_K^2,m_K^2,m_K^2,m_\eta^2,m_\pi^2,s,m_\pi^2) \, \Big(
          + 3/2\,m_\pi^2\,s^2
          - 6\,m_\pi^4\,s
          \Big)
\nonumber\\&&
       + V_{11}(m_\eta^2,m_\eta^2,m_K^2,m_K^2,m_\pi^2,s,m_\pi^2) \, \Big(
          + 1/2\,m_\pi^4\,s
          - 2\,m_\pi^6
          \Big)
\nonumber\\&&
       + V_{13}(m_K^2,m_K^2,m_\pi^2,m_K^2,m_\pi^2,s,m_\pi^2) \, \Big(
          + 21/4\,m_\pi^2\,s^2
          - 9\,m_\pi^4\,s
          - 3/4\,s^3
          \Big)
\nonumber\\&&
       + V_{13}(m_K^2,m_K^2,m_K^2,m_\eta^2,m_\pi^2,s,m_\pi^2) \, \Big(
          + 15/4\,m_\pi^2\,s^2
          - 6\,m_\pi^4\,s
          - 9/16\,s^3
          \Big)
\nonumber\\&&
       + V_{14}(m_K^2,m_K^2,m_\pi^2,m_K^2,m_\pi^2,s,m_\pi^2) \, \Big(
          + 6\,m_\pi^2\,s^2
          - 3/2\,s^3
          \Big)
\nonumber\\&&
       + V_{14}(m_K^2,m_K^2,m_K^2,m_\eta^2,m_\pi^2,s,m_\pi^2) \, \Big(
          + 9/2\,m_\pi^2\,s^2
          - 9/8\,s^3
          \Big)
\nonumber\\&&
       + V_{22}(m_\pi^2,m_\pi^2,m_\pi^2,m_\pi^2,m_\pi^2,s,m_\pi^2) \, \Big(
          - 4\,m_\pi^2\,s^2
          + 18\,m_\pi^4\,s
          - 8\,m_\pi^6
          \Big)
\nonumber\\&&
       + V_{22}(m_\pi^2,m_\pi^2,m_K^2,m_K^2,m_\pi^2,s,m_\pi^2) \, \Big(
          + 1/2\,m_\pi^2\,s^2
          - 9/4\,m_\pi^4\,s
          + m_\pi^6
          \Big)
\nonumber\\&&
       + V_{22}(m_K^2,m_K^2,m_\pi^2,m_K^2,m_\pi^2,s,m_\pi^2) \, \Big(
          - 3\,m_\pi^2\,s^2
          + 12\,m_\pi^4\,s
          \Big)
\nonumber\\&&
       + V_{22}(m_K^2,m_K^2,m_K^2,m_\eta^2,m_\pi^2,s,m_\pi^2) \, \Big(
          - 9/8\,m_\pi^2\,s^2
          + 9/2\,m_\pi^4\,s
          \Big)
\nonumber\\&&
       + V_{22}(m_\eta^2,m_\eta^2,m_K^2,m_K^2,m_\pi^2,s,m_\pi^2) \, \Big(
          - 3/4\,m_\pi^4\,s
          + 3\,m_\pi^6
          \Big)
\nonumber\\&&
       + V_{23}(m_\pi^2,m_\pi^2,m_\pi^2,m_\pi^2,m_\pi^2,s,m_\pi^2) \, \Big(
          - 18\,m_\pi^2\,s^2
          + 8\,m_\pi^4\,s
          + 4\,s^3
          \Big)
\nonumber\\&&
       + V_{23}(m_\pi^2,m_\pi^2,m_K^2,m_K^2,m_\pi^2,s,m_\pi^2) \, \Big(
          + 9/4\,m_\pi^2\,s^2
          - m_\pi^4\,s
          - 1/2\,s^3
          \Big)
\nonumber\\&&
       + V_{23}(m_K^2,m_K^2,m_\pi^2,m_K^2,m_\pi^2,s,m_\pi^2) \, \Big(
          - 12\,m_\pi^2\,s^2
          + 3\,s^3
          \Big)
\nonumber\\&&
       + V_{23}(m_K^2,m_K^2,m_K^2,m_\eta^2,m_\pi^2,s,m_\pi^2) \, \Big(
          - 9/2\,m_\pi^2\,s^2
          + 9/8\,s^3
          \Big)
\nonumber\\&&
       + V_{23}(m_\eta^2,m_\eta^2,m_K^2,m_K^2,m_\pi^2,s,m_\pi^2) \, \Big(
          + 3/4\,m_\pi^2\,s^2
          - 3\,m_\pi^4\,s
          \Big)
\nonumber\\&&
       + V_{26}(m_K^2,m_K^2,m_\pi^2,m_K^2,m_\pi^2,s,m_\pi^2) \, \Big(
          - 6\,m_\pi^2\,s^2
          + 12\,m_\pi^4\,s
          + 3/4\,s^3
          \Big)
\nonumber\\&&
       + V_{26}(m_K^2,m_K^2,m_K^2,m_\eta^2,m_\pi^2,s,m_\pi^2) \, \Big(
          - 9/2\,m_\pi^2\,s^2
          + 9\,m_\pi^4\,s
          + 9/16\,s^3
          \Big)
\nonumber\\&&
       + V_{27}(m_K^2,m_K^2,m_\pi^2,m_K^2,m_\pi^2,s,m_\pi^2) \, \Big(
          - 12\,m_\pi^2\,s^2
          + 3\,s^3
          \Big)
\nonumber\\&&
       + V_{27}(m_K^2,m_K^2,m_K^2,m_\eta^2,m_\pi^2,s,m_\pi^2) \, \Big(
          - 9\,m_\pi^2\,s^2
          + 9/4\,s^3
          \Big)
\nonumber\\&&
       + V_{29}(m_K^2,m_K^2,m_\pi^2,m_K^2,m_\pi^2,s,m_\pi^2) \, \Big(
          - 6\,m_\pi^2\,s^2
          + 3/2\,s^3
          \Big)
\nonumber\\&&
       + V_{29}(m_K^2,m_K^2,m_K^2,m_\eta^2,m_\pi^2,s,m_\pi^2) \, \Big(
          - 9/2\,m_\pi^2\,s^2
          + 9/8\,s^3
          \Big)
\nonumber\\&&
       + V_{210}(m_\pi^2,m_\pi^2,m_\pi^2,m_\pi^2,m_\pi^2,s,m_\pi^2) \, \Big(
          + 40\,m_\pi^2\,s
          - 16\,m_\pi^4
          - 16\,s^2
          \Big)
\nonumber\\&&
       + V_{210}(m_\pi^2,m_\pi^2,m_K^2,m_K^2,m_\pi^2,s,m_\pi^2) \, \Big(
          + 20\,m_\pi^2\,s
          - 8\,m_\pi^4
          - 8\,s^2
          \Big)
\nonumber\\&&
       + V_{210}(m_K^2,m_K^2,m_\pi^2,m_K^2,m_\pi^2,s,m_\pi^2) \, \Big(
          + 9/2\,m_\pi^2\,s
          - 9/4\,s^2
          \Big)
\nonumber\\&&
       + V_{210}(m_K^2,m_K^2,m_K^2,m_\eta^2,m_\pi^2,s,m_\pi^2) \, \Big(
          + 9/2\,m_\pi^2\,s
          - 9/4\,s^2
          \Big)
\nonumber\\&&
       + V_{211}(m_\pi^2,m_\pi^2,m_\pi^2,m_\pi^2,m_\pi^2,s,m_\pi^2) \, \Big(
          - 16\,m_\pi^2\,s^2
          + 40\,m_\pi^4\,s
          - 16\,m_\pi^6
          \Big)
\nonumber\\&&
       + V_{211}(m_\pi^2,m_\pi^2,m_K^2,m_K^2,m_\pi^2,s,m_\pi^2) \, \Big(
          - 8\,m_\pi^2\,s^2
          + 20\,m_\pi^4\,s
          - 8\,m_\pi^6
          \Big)
\nonumber\\&&
       + V_{211}(m_K^2,m_K^2,m_\pi^2,m_K^2,m_\pi^2,s,m_\pi^2) \, \Big(
          - 9/4\,m_\pi^2\,s^2
          + 9/2\,m_\pi^4\,s
          \Big)
\nonumber\\&&
       + V_{211}(m_K^2,m_K^2,m_K^2,m_\eta^2,m_\pi^2,s,m_\pi^2) \, \Big(
          - 9/4\,m_\pi^2\,s^2
          + 9/2\,m_\pi^4\,s
          \Big)
\nonumber\\&&
       + V_{212}(m_\pi^2,m_\pi^2,m_\pi^2,m_\pi^2,m_\pi^2,s,m_\pi^2) \, \Big(
          + 4\,m_\pi^2\,s^2
          - 8\,s^3
          \Big)
\nonumber\\&&
       + V_{212}(m_\pi^2,m_\pi^2,m_K^2,m_K^2,m_\pi^2,s,m_\pi^2) \, \Big(
          + 2\,m_\pi^2\,s^2
          - 4\,s^3
          \Big)
\nonumber\\&&
       + V_{212}(m_K^2,m_K^2,m_\pi^2,m_K^2,m_\pi^2,s,m_\pi^2) \, \Big(
          - 9/8\,s^3
          \Big)
\nonumber\\&&
       + V_{212}(m_K^2,m_K^2,m_K^2,m_\eta^2,m_\pi^2,s,m_\pi^2) \, \Big(
          - 9/8\,s^3
          \Big)
\nonumber\\&&
       + V_{213}(m_\pi^2,m_\pi^2,m_\pi^2,m_\pi^2,m_\pi^2,s,m_\pi^2) \, \Big(
          + 4\,m_\pi^2\,s^2
          - 8\,s^3
          \Big)
\nonumber\\&&
       + V_{213}(m_\pi^2,m_\pi^2,m_K^2,m_K^2,m_\pi^2,s,m_\pi^2) \, \Big(
          + 2\,m_\pi^2\,s^2
          - 4\,s^3
          \Big)
\nonumber\\&&
       + V_{213}(m_K^2,m_K^2,m_\pi^2,m_K^2,m_\pi^2,s,m_\pi^2) \, \Big(
          - 9/8\,s^3
          \Big)
\nonumber\\&&
       + V_{213}(m_K^2,m_K^2,m_K^2,m_\eta^2,m_\pi^2,s,m_\pi^2) \, \Big(
          - 9/8\,s^3
          \Big)\,.
\ea

\ba
V^1_{\mathbf{L}}(s) &=&
       + \overline{B}(m_\pi^2,m_\pi^2,s) \, \Big(
          + 32/9\,m_\pi^2\,L^r_{1}s
          - 16/9\,m_\pi^2\,L^r_{2}s
          + 16/9\,m_\pi^2\,L^r_{3}s
          + 16/9\,m_\pi^2\,L^r_{4}s
\nonumber\\&&
          + 8/9\,m_\pi^2\,L^r_{5}s
          - 64/9\,m_\pi^4\,L^r_{4}
          - 32/9\,m_\pi^4\,L^r_{5}
          - 8/9\,L^r_{1}s^2
          + 4/9\,L^r_{2}s^2
          - 4/9\,L^r_{3}s^2
          \Big)
\nonumber\\&&
       + \overline{B}(m_K^2,m_K^2,s) \, \Big(
          - 16/9\,m_\pi^2\,m_K^2\,L^r_{5}
          + 4/9\,m_\pi^2\,L^r_{5}s
          + 8/9\,m_K^2\,L^r_{3}s
          - 2/9\,L^r_{3}s^2
          \Big)\,. 
\nonumber\\&&
\ea

\ba
V^1_{\mathbf{6}}(s) &=&
       + \overline{B}(m_\pi^2,m_\pi^2,s)\,\frac{1}{16\pi^2} \, \Big(
          + 4/27\,m_\pi^2\,m_K^2
          - 11/27\,m_\pi^2\,s
          + 28/27\,m_\pi^4
          - 1/27\,m_K^2\,s
\nonumber\\&&
          + 1/27\,s^2
          \Big)
\nonumber\\&&
       + \overline{B}(m_\pi^2,m_\pi^2,s)^2 \, \Big(
          - 2/27\,m_\pi^2\,s
          + 4/27\,m_\pi^4
          + 1/108\,s^2
          \Big)
\nonumber\\&&
       + \overline{B}(m_\pi^2,m_\pi^2,s)\,\overline{B}(m_K^2,m_K^2,s) \, \Big(
          + 4/27\,m_\pi^2\,m_K^2
          - 1/27\,m_\pi^2\,s
          - 1/27\,m_K^2\,s
\nonumber\\&&
          + 1/108\,s^2
          \Big)
\nonumber\\&&
       + \overline{B}(m_\pi^2,m_\pi^2,s) \, \Big(
          - 8/27\,m_\pi^2\,\overline{A}(m_\pi^2)
          - 4/27\,m_\pi^2\,\overline{A}(m_K^2)
          + 2/27\,s\,\overline{A}(m_\pi^2)
\nonumber\\&&
          + 1/27\,s\,\overline{A}(m_K^2)
          \Big)
\nonumber\\&&
       + \overline{B}(m_K^2,m_K^2,s)\,\frac{1}{16\pi^2} \, \Big(
          + 10/27\,m_\pi^2\,m_K^2
          - 5/54\,m_\pi^2\,s
          - 5/54\,m_K^2\,s
          + 2/27\,m_K^4
\nonumber\\&&
          + 1/54\,s^2
          \Big)
\nonumber\\&&
       + \overline{B}(m_K^2,m_K^2,s)^2 \, \Big(
          - 1/54\,m_K^2\,s
          + 1/27\,m_K^4
          + 1/432\,s^2
          \Big)
\nonumber\\&&
       + \overline{B}(m_K^2,m_K^2,s) \, \Big(
          - 4/27\,m_K^2\,\overline{A}(m_\pi^2)
          - 2/27\,m_K^2\,\overline{A}(m_K^2)
          + 1/27\,s\,\overline{A}(m_\pi^2)
\nonumber\\&&
          + 1/54\,s\,\overline{A}(m_K^2)
          \Big)
\nonumber\\&&
       + \overline{B}^\epsilon(m_\pi^2,m_\pi^2,s)\,\frac{1}{16\pi^2} \, \Big(
          - 7/18\,m_\pi^2\,s
          + 10/9\,m_\pi^4
          + 1/36\,s^2
          \Big)
\nonumber\\&&
       + \overline{B}^\epsilon(m_K^2,m_K^2,s)\,\frac{1}{16\pi^2} \, \Big(
          + 1/3\,m_\pi^2\,m_K^2
          - 1/12\,m_\pi^2\,s
          - 1/18\,m_K^2\,s
          + 1/72\,s^2
          \Big)\,. \nonumber\\&&
\ea

\ba
V^1_{\mathbf{V}}(s)&=&
       + V_{0}(m_\pi^2,m_\pi^2,m_\pi^2,m_\pi^2,m_\pi^2,s,m_\pi^2) \, \Big(
          - 1/3\,m_\pi^2\,s
          \Big)
\nonumber\\&&
       + V_{0}(m_K^2,m_K^2,m_\pi^2,m_K^2,m_\pi^2,s,m_\pi^2) \, \Big(
          - 2/3\,m_\pi^2\,s
          + 1/6\,s^2
          \Big)
\nonumber\\&&
       + V_{0}(m_K^2,m_K^2,m_K^2,m_\eta^2,m_\pi^2,s,m_\pi^2) \, \Big(
          - 1/3\,m_\pi^2\,s
          + 1/8\,s^2
          \Big)
\nonumber\\&&
       + V_{11}(m_\pi^2,m_\pi^2,m_\pi^2,m_\pi^2,m_\pi^2,s,m_\pi^2) \, \Big(
          + 1/3\,m_\pi^2\,s
          + 1/3\,m_\pi^4
          \Big)
\nonumber\\&&
       + V_{11}(m_\pi^2,m_\pi^2,m_\eta^2,m_\eta^2,m_\pi^2,s,m_\pi^2) \, \Big(
          + 1/27\,m_\pi^4
          \Big)
\nonumber\\&&
       + V_{11}(m_K^2,m_K^2,m_\pi^2,m_K^2,m_\pi^2,s,m_\pi^2) \, \Big(
          + 2/3\,m_\pi^2\,s
          + m_\pi^4
          - 1/4\,s^2
          \Big)
\nonumber\\&&
       + V_{11}(m_K^2,m_K^2,m_K^2,m_\eta^2,m_\pi^2,s,m_\pi^2) \, \Big(
          + 1/3\,m_\pi^2\,s
          + 4/9\,m_\pi^4
          - 3/16\,s^2
          \Big)
\nonumber\\&&
       + V_{13}(m_K^2,m_K^2,m_\pi^2,m_K^2,m_\pi^2,s,m_\pi^2) \, \Big(
          + 2/3\,m_\pi^2\,s
          - 1/6\,s^2
          \Big)
\nonumber\\&&
       + V_{13}(m_K^2,m_K^2,m_K^2,m_\eta^2,m_\pi^2,s,m_\pi^2) \, \Big(
          + 1/2\,m_\pi^2\,s
          - 1/8\,s^2
          \Big)
\nonumber\\&&
       + V_{14}(m_K^2,m_K^2,m_\pi^2,m_K^2,m_\pi^2,s,m_\pi^2) \, \Big(
          - m_\pi^2\,s
          + 1/3\,s^2
          \Big)
\nonumber\\&&
       + V_{14}(m_K^2,m_K^2,m_K^2,m_\eta^2,m_\pi^2,s,m_\pi^2) \, \Big(
          - 2/3\,m_\pi^2\,s
          + 1/4\,s^2
          \Big)
\nonumber\\&&
       + V_{22}(m_\pi^2,m_\pi^2,m_\pi^2,m_\pi^2,m_\pi^2,s,m_\pi^2) \, \Big(
          - 4/3\,m_\pi^4
          \Big)
\nonumber\\&&
       + V_{22}(m_K^2,m_K^2,m_\pi^2,m_K^2,m_\pi^2,s,m_\pi^2) \, \Big(
          + 2/3\,m_\pi^2\,s
          - 8/3\,m_\pi^4
          \Big)
\nonumber\\&&
       + V_{22}(m_K^2,m_K^2,m_K^2,m_\eta^2,m_\pi^2,s,m_\pi^2) \, \Big(
          + 1/2\,m_\pi^2\,s
          - 4/3\,m_\pi^4
          \Big)
\nonumber\\&&
       + V_{23}(m_\pi^2,m_\pi^2,m_\pi^2,m_\pi^2,m_\pi^2,s,m_\pi^2) \, \Big(
          + 4/3\,m_\pi^2\,s
          \Big)
\nonumber\\&&
       + V_{23}(m_K^2,m_K^2,m_\pi^2,m_K^2,m_\pi^2,s,m_\pi^2) \, \Big(
          + 8/3\,m_\pi^2\,s
          - 2/3\,s^2
          \Big)
\nonumber\\&&
       + V_{23}(m_K^2,m_K^2,m_K^2,m_\eta^2,m_\pi^2,s,m_\pi^2) \, \Big(
          + 4/3\,m_\pi^2\,s
          - 1/2\,s^2
          \Big)
\nonumber\\&&
       + V_{26}(m_K^2,m_K^2,m_\pi^2,m_K^2,m_\pi^2,s,m_\pi^2) \, \Big(
          - 1/6\,m_\pi^2\,s
          - 2\,m_\pi^4
          + 1/6\,s^2
          \Big)
\nonumber\\&&
       + V_{26}(m_K^2,m_K^2,m_K^2,m_\eta^2,m_\pi^2,s,m_\pi^2) \, \Big(
          - 1/6\,m_\pi^2\,s
          - 4/3\,m_\pi^4
          + 1/8\,s^2
          \Big)
\nonumber\\&&
       + V_{27}(m_K^2,m_K^2,m_\pi^2,m_K^2,m_\pi^2,s,m_\pi^2) \, \Big(
          + 2\,m_\pi^2\,s
          - 2/3\,s^2
          \Big)
\nonumber\\&&
       + V_{27}(m_K^2,m_K^2,m_K^2,m_\eta^2,m_\pi^2,s,m_\pi^2) \, \Big(
          + 4/3\,m_\pi^2\,s
          - 1/2\,s^2
          \Big)
\nonumber\\&&
       + V_{29}(m_K^2,m_K^2,m_\pi^2,m_K^2,m_\pi^2,s,m_\pi^2) \, \Big(
          + 11/3\,m_\pi^2\,s
          - s^2
          \Big)
\nonumber\\&&
       + V_{29}(m_K^2,m_K^2,m_K^2,m_\eta^2,m_\pi^2,s,m_\pi^2) \, \Big(
          + 8/3\,m_\pi^2\,s
          - 3/4\,s^2
          \Big)
\nonumber\\&&
       + V_{31}(m_K^2,m_K^2,m_\pi^2,m_K^2,m_\pi^2,s,m_\pi^2) \, \Big(
          + 4\,m_\pi^2
          - s
          \Big)
\nonumber\\&&
       + V_{31}(m_K^2,m_K^2,m_K^2,m_\eta^2,m_\pi^2,s,m_\pi^2) \, \Big(
          + 3\,m_\pi^2
          - 3/4\,s
          \Big)
\nonumber\\&&
       + V_{33}(m_K^2,m_K^2,m_\pi^2,m_K^2,m_\pi^2,s,m_\pi^2) \, \Big(
          - 2/3\,m_\pi^2\,s
          + 4/3\,m_\pi^4
          + 1/12\,s^2
          \Big)
\nonumber\\&&
       + V_{33}(m_K^2,m_K^2,m_K^2,m_\eta^2,m_\pi^2,s,m_\pi^2) \, \Big(
          - 1/2\,m_\pi^2\,s
          + m_\pi^4
          + 1/16\,s^2
          \Big)
\nonumber\\&&
       + V_{213}(m_\pi^2,m_\pi^2,m_\pi^2,m_\pi^2,m_\pi^2,s,m_\pi^2) \, \Big(
          + 8/3\,m_\pi^2\,s
          - 2/3\,s^2
          \Big)
\nonumber\\&&
       + V_{213}(m_\pi^2,m_\pi^2,m_K^2,m_K^2,m_\pi^2,s,m_\pi^2) \, \Big(
          + 4/3\,m_\pi^2\,s
          - 1/3\,s^2
          \Big)
\nonumber\\&&
       + V_{213}(m_K^2,m_K^2,m_\pi^2,m_K^2,m_\pi^2,s,m_\pi^2) \, \Big(
          + m_\pi^2\,s
          - 1/4\,s^2
          \Big)
\nonumber\\&&
       + V_{213}(m_K^2,m_K^2,m_K^2,m_\eta^2,m_\pi^2,s,m_\pi^2) \, \Big(
          + m_\pi^2\,s
          - 1/4\,s^2
          \Big)
\nonumber\\&&
       + V_{311}(m_K^2,m_K^2,m_\pi^2,m_K^2,m_\pi^2,s,m_\pi^2) \, \Big(
          - 2/3\,m_\pi^2\,s
          + 8/3\,m_\pi^4
          \Big)
\nonumber\\&&
       + V_{311}(m_K^2,m_K^2,m_K^2,m_\eta^2,m_\pi^2,s,m_\pi^2) \, \Big(
          - 1/2\,m_\pi^2\,s
          + 2\,m_\pi^4
          \Big)
\nonumber\\&&
       + V_{313}(m_K^2,m_K^2,m_\pi^2,m_K^2,m_\pi^2,s,m_\pi^2) \, \Big(
          - 8/3\,m_\pi^2\,s
          + 2/3\,s^2
          \Big)
\nonumber\\&&
       + V_{313}(m_K^2,m_K^2,m_K^2,m_\eta^2,m_\pi^2,s,m_\pi^2) \, \Big(
          - 2\,m_\pi^2\,s
          + 1/2\,s^2
          \Big)
\nonumber\\&&
       + V_{314}(m_K^2,m_K^2,m_\pi^2,m_K^2,m_\pi^2,s,m_\pi^2) \, \Big(
          - 8/3\,m_\pi^2\,s
          + 2/3\,s^2
          \Big)
\nonumber\\&&
       + V_{314}(m_K^2,m_K^2,m_K^2,m_\eta^2,m_\pi^2,s,m_\pi^2) \, \Big(
          - 2\,m_\pi^2\,s
          + 1/2\,s^2
          \Big)
\nonumber\\&&
       + V_{316}(m_K^2,m_K^2,m_\pi^2,m_K^2,m_\pi^2,s,m_\pi^2) \, \Big(
          - 16/3\,m_\pi^2\,s
          + 4/3\,s^2
          \Big)
\nonumber\\&&
       + V_{316}(m_K^2,m_K^2,m_K^2,m_\eta^2,m_\pi^2,s,m_\pi^2) \, \Big(
          - 4\,m_\pi^2\,s
          + s^2
          \Big)
\nonumber\\&&
       + V_{317}(m_\pi^2,m_\pi^2,m_\pi^2,m_\pi^2,m_\pi^2,s,m_\pi^2) \, \Big(
          + 8/3\,m_\pi^2
          - 4/3\,s
          \Big)
\nonumber\\&&
       + V_{317}(m_\pi^2,m_\pi^2,m_K^2,m_K^2,m_\pi^2,s,m_\pi^2) \, \Big(
          + 4/3\,m_\pi^2
          - 2/3\,s
          \Big)
\nonumber\\&&
       + V_{317}(m_K^2,m_K^2,m_\pi^2,m_K^2,m_\pi^2,s,m_\pi^2) \, \Big(
          + m_\pi^2
          - 1/2\,s
          \Big)
\nonumber\\&&
       + V_{317}(m_K^2,m_K^2,m_K^2,m_\eta^2,m_\pi^2,s,m_\pi^2) \, \Big(
          + m_\pi^2
          - 1/2\,s
          \Big)
\nonumber\\&&
       + V_{321}(m_\pi^2,m_\pi^2,m_\pi^2,m_\pi^2,m_\pi^2,s,m_\pi^2) \, \Big(
          - 4/3\,m_\pi^2\,s
          + 8/3\,m_\pi^4
          \Big)
\nonumber\\&&
       + V_{321}(m_\pi^2,m_\pi^2,m_K^2,m_K^2,m_\pi^2,s,m_\pi^2) \, \Big(
          - 2/3\,m_\pi^2\,s
          + 4/3\,m_\pi^4
          \Big)
\nonumber\\&&
       + V_{321}(m_K^2,m_K^2,m_\pi^2,m_K^2,m_\pi^2,s,m_\pi^2) \, \Big(
          - 1/2\,m_\pi^2\,s
          + m_\pi^4
          \Big)
\nonumber\\&&
       + V_{321}(m_K^2,m_K^2,m_K^2,m_\eta^2,m_\pi^2,s,m_\pi^2) \, \Big(
          - 1/2\,m_\pi^2\,s
          + m_\pi^4
          \Big)
\nonumber\\&&
       + V_{323}(m_\pi^2,m_\pi^2,m_\pi^2,m_\pi^2,m_\pi^2,s,m_\pi^2) \, \Big(
          - 2/3\,s^2
          \Big)
\nonumber\\&&
       + V_{323}(m_\pi^2,m_\pi^2,m_K^2,m_K^2,m_\pi^2,s,m_\pi^2) \, \Big(
          - 1/3\,s^2
          \Big)
\nonumber\\&&
       + V_{323}(m_K^2,m_K^2,m_\pi^2,m_K^2,m_\pi^2,s,m_\pi^2) \, \Big(
          - 1/4\,s^2
          \Big)
\nonumber\\&&
       + V_{323}(m_K^2,m_K^2,m_K^2,m_\eta^2,m_\pi^2,s,m_\pi^2) \, \Big(
          - 1/4\,s^2
          \Big)
\nonumber\\&&
       + V_{325}(m_\pi^2,m_\pi^2,m_\pi^2,m_\pi^2,m_\pi^2,s,m_\pi^2) \, \Big(
          - 8/3\,m_\pi^2\,s
          \Big)
\nonumber\\&&
       + V_{325}(m_\pi^2,m_\pi^2,m_K^2,m_K^2,m_\pi^2,s,m_\pi^2) \, \Big(
          - 4/3\,m_\pi^2\,s
          \Big)
\nonumber\\&&
       + V_{325}(m_K^2,m_K^2,m_\pi^2,m_K^2,m_\pi^2,s,m_\pi^2) \, \Big(
          - m_\pi^2\,s
          \Big)
\nonumber\\&&
       + V_{325}(m_K^2,m_K^2,m_K^2,m_\eta^2,m_\pi^2,s,m_\pi^2) \, \Big(
          - m_\pi^2\,s
          \Big)
\nonumber\\&&
       + V_{326}(m_\pi^2,m_\pi^2,m_\pi^2,m_\pi^2,m_\pi^2,s,m_\pi^2) \, \Big(
          - 16/3\,m_\pi^2\,s
          + 4/3\,s^2
          \Big)
\nonumber\\&&
       + V_{326}(m_\pi^2,m_\pi^2,m_K^2,m_K^2,m_\pi^2,s,m_\pi^2) \, \Big(
          - 8/3\,m_\pi^2\,s
          + 2/3\,s^2
          \Big)
\nonumber\\&&
       + V_{326}(m_K^2,m_K^2,m_\pi^2,m_K^2,m_\pi^2,s,m_\pi^2) \, \Big(
          - 2\,m_\pi^2\,s
          + 1/2\,s^2
          \Big)
\nonumber\\&&
       + V_{326}(m_K^2,m_K^2,m_K^2,m_\eta^2,m_\pi^2,s,m_\pi^2) \, \Big(
          - 2\,m_\pi^2\,s
          + 1/2\,s^2
          \Big)
\,.
\ea

\ba
V^2_{\mathbf{L}}(s)&=&
       + \overline{B}(m_\pi^2,m_\pi^2,s) \, \Big(
          + 64/3\,m_\pi^2\,L^r_{1}s^2
          + 176/3\,m_\pi^2\,L^r_{2}s^2
          + 32/3\,m_\pi^2\,L^r_{3}s^2
\nonumber\\&&
          + 16\,m_\pi^2\,L^r_{4}s^2
          + 8\,m_\pi^2\,L^r_{5}s^2
          - 128/3\,m_\pi^4\,L^r_{1}s
          - 352/3\,m_\pi^4\,L^r_{2}s
          - 64/3\,m_\pi^4\,L^r_{3}s
\nonumber\\&&
          - 32\,m_\pi^4\,L^r_{4}s
          - 16\,m_\pi^4\,L^r_{5}s
          - 64\,m_\pi^4\,L^r_{6}s
          - 32\,m_\pi^4\,L^r_{8}s
          + 128/3\,m_\pi^6\,L^r_{1}
\nonumber\\&&
          + 256/3\,m_\pi^6\,L^r_{2}
          + 64/3\,m_\pi^6\,L^r_{3}
          + 128\,m_\pi^6\,L^r_{6}
          + 64\,m_\pi^6\,L^r_{8}
          - 16/3\,L^r_{1}s^3
\nonumber\\&&
          - 32/3\,L^r_{2}s^3
          - 8/3\,L^r_{3}s^3
          \Big)\,.
\ea

\ba
V^2_{\mathbf{6}}(s)&=&
       + \overline{B}(m_\pi^2,m_\pi^2,s)\,\frac{1}{16\pi^2} \, \Big(
          - 5/6\,m_\pi^2\,s^2
          + 8/3\,m_\pi^4\,s
          - 8/3\,m_\pi^6
          + 1/12\,s^3
          \Big)
\nonumber\\&&
       + \overline{B}(m_\pi^2,m_\pi^2,s)^2 \, \Big(
          + 3/2\,m_\pi^2\,s^2
          - 3\,m_\pi^4\,s
          + 2\,m_\pi^6
          - 1/4\,s^3
          \Big)
\nonumber\\&&
       + \overline{B}(m_\pi^2,m_\pi^2,s) \, \Big(
          - 8\,m_\pi^2\,s\,\overline{A}(m_\pi^2)
          - 4\,m_\pi^2\,s\,\overline{A}(m_K^2)
          + 8\,m_\pi^4\,\overline{A}(m_\pi^2)
\nonumber\\&&
          + 4\,m_\pi^4\,\overline{A}(m_K^2)
          + 2\,s^2\,\overline{A}(m_\pi^2)
          + s^2\,\overline{A}(m_K^2)
          \Big)
\nonumber\\&&
       + \overline{B}^\epsilon(m_\pi^2,m_\pi^2,s)\,\frac{1}{16\pi^2} \, \Big(
          - 4\,m_\pi^2\,m_K^2\,s
          - 4\,m_\pi^2\,s^2
          + 4\,m_\pi^4\,m_K^2
          + 73/9\,m_\pi^4\,s
\nonumber\\&&
          - 56/9\,m_\pi^6
          + m_K^2\,s^2
          + 3/4\,s^3
          \Big)\,.
\ea

\ba
V^2_{\mathbf{V}}(s) &=&
       + V_{0}(m_\pi^2,m_\pi^2,m_\pi^2,m_\pi^2,m_\pi^2,s,m_\pi^2) \, \Big(
          + m_\pi^2\,s^2
          - 7\,m_\pi^4\,s
          + 10\,m_\pi^6
          \Big)
\nonumber\\&&
       + V_{0}(m_\pi^2,m_\pi^2,m_K^2,m_K^2,m_\pi^2,s,m_\pi^2) \, \Big(
          + 1/2\,m_\pi^2\,s^2
          - 3\,m_\pi^4\,s
          + 4\,m_\pi^6
          \Big)
\nonumber\\&&
       + V_{0}(m_\pi^2,m_\pi^2,m_\eta^2,m_\eta^2,m_\pi^2,s,m_\pi^2) \, \Big(
          - 1/9\,m_\pi^4\,s
          + 2/9\,m_\pi^6
          \Big)
\nonumber\\&&
       + V_{11}(m_\pi^2,m_\pi^2,m_\pi^2,m_\pi^2,m_\pi^2,s,m_\pi^2) \, \Big(
          - 3\,m_\pi^2\,s^2
          + 18\,m_\pi^4\,s
          - 24\,m_\pi^6
          \Big)
\nonumber\\&&
       + V_{11}(m_\pi^2,m_\pi^2,m_K^2,m_K^2,m_\pi^2,s,m_\pi^2) \, \Big(
          - m_\pi^2\,s^2
          + 6\,m_\pi^4\,s
          - 8\,m_\pi^6
          \Big)
\nonumber\\&&
       + V_{22}(m_\pi^2,m_\pi^2,m_\pi^2,m_\pi^2,m_\pi^2,s,m_\pi^2) \, \Big(
          + 2\,m_\pi^2\,s^2
          - 12\,m_\pi^4\,s
          + 16\,m_\pi^6
          \Big)
\nonumber\\&&
       + V_{22}(m_\pi^2,m_\pi^2,m_K^2,m_K^2,m_\pi^2,s,m_\pi^2) \, \Big(
          + 1/2\,m_\pi^2\,s^2
          - 3\,m_\pi^4\,s
          + 4\,m_\pi^6
          \Big)
\nonumber\\&&
       + V_{23}(m_\pi^2,m_\pi^2,m_\pi^2,m_\pi^2,m_\pi^2,s,m_\pi^2) \, \Big(
          + 12\,m_\pi^2\,s^2
          - 16\,m_\pi^4\,s
          - 2\,s^3
          \Big)
\nonumber\\&&
       + V_{23}(m_\pi^2,m_\pi^2,m_K^2,m_K^2,m_\pi^2,s,m_\pi^2) \, \Big(
          + 3\,m_\pi^2\,s^2
          - 4\,m_\pi^4\,s
          - 1/2\,s^3
          \Big)
\nonumber\\&&
       + V_{210}(m_\pi^2,m_\pi^2,m_\pi^2,m_\pi^2,m_\pi^2,s,m_\pi^2) \, \Big(
          + 16\,m_\pi^2\,s
          - 16\,m_\pi^4
          - 4\,s^2
          \Big)
\nonumber\\&&
       + V_{210}(m_\pi^2,m_\pi^2,m_K^2,m_K^2,m_\pi^2,s,m_\pi^2) \, \Big(
          + 8\,m_\pi^2\,s
          - 8\,m_\pi^4
          - 2\,s^2
          \Big)
\nonumber\\&&
       + V_{211}(m_\pi^2,m_\pi^2,m_\pi^2,m_\pi^2,m_\pi^2,s,m_\pi^2) \, \Big(
          - 4\,m_\pi^2\,s^2
          + 16\,m_\pi^4\,s
          - 16\,m_\pi^6
          \Big)
\nonumber\\&&
       + V_{211}(m_\pi^2,m_\pi^2,m_K^2,m_K^2,m_\pi^2,s,m_\pi^2) \, \Big(
          - 2\,m_\pi^2\,s^2
          + 8\,m_\pi^4\,s
          - 8\,m_\pi^6
          \Big)
\nonumber\\&&
       + V_{212}(m_\pi^2,m_\pi^2,m_\pi^2,m_\pi^2,m_\pi^2,s,m_\pi^2) \, \Big(
          + 4\,m_\pi^2\,s^2
          - 2\,s^3
          \Big)
\nonumber\\&&
       + V_{212}(m_\pi^2,m_\pi^2,m_K^2,m_K^2,m_\pi^2,s,m_\pi^2) \, \Big(
          + 2\,m_\pi^2\,s^2
          - s^3
          \Big)
\nonumber\\&&
       + V_{213}(m_\pi^2,m_\pi^2,m_\pi^2,m_\pi^2,m_\pi^2,s,m_\pi^2) \, \Big(
          + 4\,m_\pi^2\,s^2
          - 2\,s^3
          \Big)
\nonumber\\&&
       + V_{213}(m_\pi^2,m_\pi^2,m_K^2,m_K^2,m_\pi^2,s,m_\pi^2) \, \Big(
          + 2\,m_\pi^2\,s^2
          - s^3
          \Big)\,.
\ea

\end{document}